\documentclass[showkeys,aps,twocolumn,showpacs,preprintnumbers,amsmath,amssymb,prd,letterpaper,floatfix,nofootinbib,superscriptaddress]{revtex4-1}
\usepackage{amssymb}
\usepackage{graphicx}
\usepackage{longtable}
\usepackage{hyperref}
\usepackage{braket}
\usepackage{mathbbol}
\usepackage{color}
\usepackage{xcolor}
\usepackage{cancel}
\usepackage[utf8]{inputenc}
\usepackage{soul}

\begin{document}
\def\Regensburg{Institute for Theoretical Physics, University of Regensburg, 93040 Regensburg, Germany}
\def\Mainz{Helmholtz-Institut Mainz, 55099 Mainz, Germany}
\title{Identifying spin and parity of charmonia in flight with lattice QCD}
\author{M.~Padmanath}
\email{Padmanath.M@physik.uni-regensburg.de}
\affiliation{\Regensburg}
\author{Sara~Collins}
\affiliation{\Regensburg}
\author{Daniel~Mohler}
\affiliation{\Mainz}
\affiliation{Johannes Gutenberg-Universit\"at Mainz, 55099 Mainz, Germany}
\author{Stefano Piemonte}
\affiliation{\Regensburg}
\author{Sasa~Prelovsek}
\email{sasa.prelovsek@ijs.si}
\affiliation{\Regensburg}
\affiliation{Faculty of Mathematics and Physics, University of Ljubljana, 1000 Ljubljana, Slovenia}
\affiliation{Jozef Stefan Institute, 1000 Ljubljana, Slovenia}
\author{Andreas~Sch\"{a}fer}  
\affiliation{\Regensburg}
\author{Simon~Weishaeupl}  
\affiliation{\Regensburg}

\date{\today}
\begin{abstract}
The spectrum of charmonium resonances contains a number of unanticipated states along with several conventional 
quark-model excitations. The hadrons of different quantum numbers $J^P$ appear in a fairly narrow energy band, 
where $J^P$ refers to the spin-parity of a hadron in its rest frame. This poses a challenge for Lattice QCD studies of 
(coupled-channel) meson-meson scattering aimed at the determination of scattering amplitudes and resonance pole 
positions. A wealth of information for this purpose can be obtained from the lattice spectra in frames with 
nonzero total momentum. These are particularly dense since hadrons with different $J^P$ contribute to any given 
lattice irreducible representation.  This is because $J^P$ is not  a good quantum number in flight, and also 
because the continuum symmetry is reduced  on the lattice. In this paper we address the assignment of the 
underlying continuum $J^P$ quantum numbers to charmonia in flight  using a $N_f = 2 + 1$ CLS ensemble. As a 
first step, we apply the single-hadron approach, where only interpolating fields of quark-antiquark type  
are used. The approach follows techniques previously applied to the light meson spectrum by the Hadron Spectrum 
Collaboration. The resulting  spectra of charmonia with assigned $J^P$ will provide valuable information for 
the parameterization of (resonant) amplitudes in future determinations of resonance properties with lattice QCD.
\end{abstract}
\maketitle

\preprint{MITP/18-111} 

\section{Introduction}

Over the past two decades many interesting resonance-like structures have been discovered in hadronic 
final states in the energy regime of heavy quarkonium ($\bar cc$ or $\bar bb$). A large collection of 
these structures (generally referred to as XYZs) appears close to open flavor strong decay thresholds and 
does not fit into a simple non-relativistic quark-antiquark picture. Theoretically, resonance peaks are 
associated with pole singularities of scattering amplitudes in the complex energy plane. Hence their 
nature and properties have to be inferred from the respective scattering matrices and from their decays. 
Potential models and effective field theories, that approximate certain regimes of the strong interaction, 
provide insight into describing these resonance peaks in various different ways including as tetraquarks 
(typically compact diquark-antidiquark states), mesonic molecules, hadro-quarkonia or hybrid mesons with 
excited gluonic content. There are also interpretations in which these hadronic resonance structures arise 
from strongly coupled scattering channels. For a detailed review on the different theoretical 
investigations, see Refs. \cite{Lebed:2016hpi,Esposito:2016noz}. In the end, understanding these hadron
resonances from first principles requires non-perturbative approaches such as lattice QCD. 

Lattice QCD, which is a systematically improvable {\it ab-initio} non-perturbative formulation, 
has proven to be a very powerful tool for investigating the physics of hadrons \cite{Aoki:2016frl}. Many 
lattice calculations of the excited charmonium spectrum have been performed by now, for example  
\cite{Dudek:2007wv,Bali:2011rd,Liu:2012ze,Mohler:2012na,Cheung:2016bym}. So far, only one lattice simulation studied 
the charmonium resonances above open charm threshold, performing a finite volume analysis using 
L\"uscher's method \cite{Luscher:1986pf,Luscher:1990ux,Luscher:1991cf} and obtaining 
exploratory results for masses as well as decay widths in the vector and scalar channels
\cite{Lang:2015sba}. In particular, all previous studies were performed in the rest frame, which 
provides only rather limited information on the scattering matrix. Lattice calculations of hadronic systems 
in moving frames ({\it i.e.} with non-zero total momentum) provide additional information on the relevant 
two-meson scattering matrices and have been quite successful in providing a comprehensive understanding 
of various hadron resonances (for a review, see Ref. \cite{Briceno:2017max}).


The first step towards a detailed finite volume analysis is a rigorous extraction of the excited 
hadron spectrum on the lattice in multiple inertial frames, followed by a reliable identification of 
the continuum quantum numbers. In this work, we investigate the charmonium spectrum on the 
lattice in the lowest three inertial frames with the square of the total spatial momenta, $|{\bf p}|^2=0,~1$ and 
2 (in units of $(2\pi/L)^2$, where $L$ is the spatial extent of the lattice). Compared to the rest frame, 
the determination of the spectrum on the lattice in moving frames poses additional challenges. This holds, in
particular, for the charmonium spectrum, since there are a number of states with different spin ($J$) 
and parity ($P$) in a narrow energy region and hence disentangling the $J^P$ information becomes 
non-trivial. 

The main aim of this paper is to extract the charmonium spectrum on the lattice in the rest frame and 
in moving frames and to discuss our procedure for determining the underlying $J^P$. Note that spin 
and parity refer to continuum quantum numbers of a hadron in its rest frame. Experimentally, $J^P$ for 
hadronic resonances are determined from partial-waves of their decay products transformed to their center 
of momentum frame. On the lattice, this information has to be inferred from the hadron energy spectrum 
and the overlaps with the lattice interpolating fields. This is rather challenging as lattice eigenstates transform according 
to irreducible representations (irreps) of the reduced symmetry group on the lattice. Hence infinitely 
many continuum states are constrained to appear in the spectrum of a finite number of lattice irreps. 
This means that continuum \footnote{We refer to the $a\to 0$ and $L\to \infty$ limits, when referring to 
the $J$ and $P$ of lattice states.} eigenstates with different $J$ contribute to a given irrep of the 
octahedral group ($O_h$) on a cubic lattice. For ${\bf p}\not = 0$, the symmetry is further reduced to 
the little groups, implying that the continuum eigenstates not only with different $J$ but also with 
different parity, $P$, contribute to a given irrep of the respective little group. For example, the only 
irrep of the $Dic_2$ little group (corresponding to ${\bf p}=(1,1,0)$) that contains scalar charmonia 
with $J^P=0^+$, also contains states with $J^P=1^-,~2^{\pm},~3^{\pm},...$ . Thus the charmonium spectrum 
on the lattice can be quite dense - with several states of different $J^P$ contributing to a single 
lattice irrep. A rigorous spin-identification procedure for the charmonium spectrum on the lattice 
is therefore desirable.

To identify the spin and parity, we follow the approach proposed by the Hadron Spectrum 
Collaboration (HSC) and first applied to light iso-vector mesons at rest \cite{Dudek:2010wm} and 
in flight \cite{Thomas:2011rh}. We construct correlation functions from single meson interpolators
on an $N_f = 2+1$ ensemble generated by the ``Coordinated Lattice Simulations" (CLS) consortium 
\cite{Bruno:2014jqa,Mohler:2017wnb} with $m_\pi \simeq 280$ MeV, $L\simeq2.1 ~\textrm{fm}$ and lattice 
spacing $a\simeq0.086$ fm. Using these correlation functions, we illustrate the spin identification procedure 
that allows us to determine the underlying $J^P$.
For practical reasons we limit our discussion to $J\le3$ (and consequently to helicities  
$|\lambda|\leq 3$) in all irreducible representations for the three inertial frames considered. Some 
preliminary results have been presented in Ref.~\cite{Prelovsek:2017eaa}. This spin identification procedure 
can be straightforwardly applied to excited charmonium studies on different lattices.

We restrict ourselves to the single-hadron approach employing only quark-antiquark interpolating 
fields and assuming that this procedure qualitatively captures the single-meson spectrum in the finite volume. In 
particular this approach does not aim at extracting the full spectrum of multi-hadron scattering 
states in finite volume. 
We briefly address indications for
the presence of meson-meson admixtures and postpone a rigorous extraction of scattering amplitudes from a
basis of meson-meson and quark-antiquark interpolators to future calculations. We emphasize that 
realizing this technology for multiple inertial frames within the single-hadron approach is a valuable 
first step in a precise determination of the complete finite-volume spectrum, which will then be used in 
the extraction of resonance information \cite{Thomas:2011rh}.

We are currently performing an analysis of excited charmonia 
beyond the single-hadron approach on CLS lattices, including the ensemble used here; preliminary results were 
presented in Ref. \cite{Bali:2018xkm}. This study also considers charmonium systems with non-zero momenta. 
The results from the present study can inform (model) parametrizations of scattering amplitudes with narrow 
resonances.

The layout of this paper is as follows. In Section \ref{sec:idproc}, we describe our spin identification 
procedure for charmonium at rest and in the moving frames. The construction of meson interpolators 
used in this study is detailed in Section \ref{sec:operators} and our lattice setup is briefly outlined 
in Section \ref{sec:setup}.  Section \ref{sec:results} illustrates our spin identification procedure with examples 
and our conclusions are presented in Section \ref{sec:conclusions}.

\section{Charmonia at rest and in flight on the lattice}\label{sec:idproc} 

Energy eigenvalues of charmonia, $E_n$, are extracted on the lattice by computing two-point 
correlation functions constructed from annihilation (creation) operators 
$O^{(\dagger)}_{i,\Lambda^C}$ of type $\bar cc$, inserted at initial and final times, $t'$ and $t_f$, 
respectively:
\begin{align}
  \label{eq:2-1}
  C_{ij}(t_f-t')&=\langle O_{i,\Lambda^C}(t_f)O^{\dagger}_{j,\Lambda^C}(t') \rangle
\nonumber\\&=\sum_{n=1} Z_i^{n}Z_j^{n*}e^{-E_n(t_f-t')}~.
  \end{align} 
The interpolators ($O^{\dagger}_{j,\Lambda^C}$) are chosen to transform according to the 
irreps $\Lambda^C$ of the lattice symmetry group. Subscripts
$i$ and $j$ are indices referring to the operator label. The spectral decomposition 
contains all states that transform according to the lattice irrep $\Lambda^C$. Charmonia 
are isosinglets and charge parity $C$ is a good quantum number at rest as well as in flight, 
also on the lattice. The overlap of eigenstate $n$ with an operator with label $i$ is given by 
$Z_i^n=\langle O_{i,\Lambda^C}|n\rangle$.

A tower of low lying energy levels for a given channel can be determined by employing a large 
basis of interpolators to form correlation matrices and solving the generalized eigenvalue 
problem~(GEVP)\cite{Michael:1985ne,Luscher:1985dn,Blossier:2009kd}. As states with different 
rest frame quantum numbers $J^P$ contribute to correlators corresponding to a given $\Lambda^C$ 
of the reduced symmetry on the lattice, assigning values of spin and parity to these levels 
is difficult. We overcome these difficulties by adopting an approach developed in Refs. 
\cite{Dudek:2010wm,Thomas:2011rh}, where interpolators well-suited for the GEVP,
while also aiding the assignment of spin, are constructed. 

In the following, we consider in detail the spin-parity assignments of levels with $J\le 3$, 
first for the more familiar case of ${\bf p=0}$ and then for the more challenging case of 
${\bf p\neq 0}$. In each case, we briefly discuss the symmetries and associated quantum numbers 
relevant in the continuum and on the lattice~(which determine which $J^P$ can contribute to a 
given lattice irrep). Then we summarize our spin identification procedure based on energies 
and $Z$ factors.

\subsection{Mesons at rest}\label{sec:basicsrest}

Spin $J$, its $z$-component $M$ and parity $P$ are good quantum numbers for a meson at rest 
in the infinite volume continuum.  The corresponding eigenstates are labeled as 
$|{\bf p}\!=\!\mathbf{0};J^{PC}\rangle$. On the lattice $P$ remains a good quantum number. 
Lattice eigenstates transform according to one of the five irreps of the octahedral group $O_h$, 
given in Table \ref{tab:2-1}. Group-theoretical subduction relates the infinite number of 
continuum spins to the finite number of lattice irreps as shown for $J\le 3$ in Table 
\ref{tab:2-1}.  Since several $J$ contribute to each lattice irrep $\Lambda$, 
the spin assignment of lattice energy levels within each $\Lambda^C$ is not immediate. 

Following Ref. \cite{Dudek:2010wm}, we first construct interpolators in the continuum with 
good quantum numbers $J$, $M$ and $P$ at ${\bf p}={\bf 0}$ ($O^{J^{PC},M}_{i}$). Using these 
continuum interpolators, one then builds the lattice interpolators $O^{[J^{PC}]}_{i,\Lambda^C}$ 
that transform according to irreps of $O_h$ \footnote{Lattice irreps $\Lambda$ can be 
multi-dimensional. For brevity, in this section we suppress the irrep row index without any 
loss of generality.}. For details, see Section \ref{sec:operators}. In this way these lattice 
interpolators are expected to have good overlaps with states of the respective continuum quantum 
numbers, aiding their clean determination. Below we outline the `guidelines' we follow in our 
spin assignment procedure for mesons at rest:

\begin{itemize}

\item {\bf Trivial assignments}: According to Table \ref{tab:2-1}, continuum states 
with $J=0$ are expected to appear in the $A_1$ irrep. Assuming negligible contributions from any 
$J\ge4$ spin state in the energy range considered, all extracted levels in the $A_1$ irrep can be
trivially assigned $J=0$. Analogously, the levels in $A_2$ and $E$ can only result from $J=3$ and 
$J=2$ continuum states, respectively.

\item {\bf (Non-)degeneracy of energy levels}: The appearance of a set of near degenerate energy 
levels (up to discretization and finite volume effects) or the absence of any near degenerate partner 
level across different lattice irreps immediately suggests possible spin assignments. For example, 
the appearance of a level in the $T_1$ irrep of $O_h$ with no near degenerate partner levels in 
any other irreps suggests a $J=1$ spin assignment ({\it c.f.} Table \ref{tab:2-1}). Similarly, three 
near degenerate energy levels, one each in the spectrum of the $T_1$, $T_2$ and $A_2$ irreps of $O_h$, 
suggest a $J=3$ assignment.

\item {\bf Enhanced $Z$ factors}: Those lattice eigenstates resembling continuum states with quantum 
numbers $J^{PC}$ are expected to have the largest overlap factors with the lattice operators, 
$O^{[J^{PC}]}_{i,\Lambda^C}$. Hence, enhanced values of $Z_i^n =\langle O^{[J^{PC}]}_{i,\Lambda^C}|n\rangle$ 
suggest a spin-parity assignment $J^P$. To make such an assessment, we utilize the quantity 
$\tilde Z_i^n$, which is normalized to unity with respect to the largest $Z_i^n$ for a given 
operator with label $i$ across all lattice states $n$ \cite{Dudek:2010wm}\footnote{Due to the chosen 
normalization, the $\tilde Z$ factors might be  artificially large for some operators $O^{[J^{PC}]}_{i,\Lambda} $ 
if there is no state with this $J^{PC}$ among the eigenstates in irrep $\Lambda$. The $Z$ factor for such 
an operator is generally smaller than other $Z$ factors and also gives valuable information. So it is
helpful to monitor $Z$ in addition to $\tilde Z$ in all cases.},
{\it i.e.} 
\begin{equation}
\tilde Z_i^n = \frac{Z_i^n}{max_n(Z_i^n)}\le 1.
\label{Ztilde}
\end{equation}
If the effects of rotational symmetry breaking are sufficiently small, one expects 
\begin{equation}
\langle O^{[J^{PC}]}_{i,\Lambda^C}|\mathbf{0},J^{PC}\rangle \gg \langle O^{[J^{PC}]}_{i,\Lambda^C}|\mathbf{0},J'^{PC}\rangle,
\label{Zenhance}
\end{equation}
for $J'\ne J$.

\item {\bf Degeneracy in $Z$ factors for spin $J>1$}: \\ One expects similar $Z$ factors for 
a continuum state $|{\bf p}\!=\!\mathbf{0};J^{PC}\rangle$ with interpolators subduced from the same 
continuum operator $O^{J^{PC}}_i$,
\begin{equation}
\langle O^{[J^{PC}]}_{i,\Lambda_1^C}|\mathbf{0},J^{PC}\rangle \simeq \langle O^{[J^{PC}]}_{i,\Lambda_2^{C}}|\mathbf{0},J^{PC}\rangle.
\label{Zequal}
\end{equation}
For example, a $J=3$ continuum state manifests as three almost degenerate energy levels, one each 
in the spectrum obtained using $T_1$, $T_2$ and $A_2$ operators. Each of these levels is expected to have 
degenerate $Z_i^n$ factors for lattice interpolators subduced from the same continuum interpolator with 
spin 3. {\it i.e.}
\begin{equation}
\langle O^{[3]}_{i,T_1}|n_{T_1}\rangle \simeq \langle O^{[3]}_{i,T_2}|n_{T_2}\rangle \simeq \langle O^{[3]}_{i,A_2}|n_{A_2}\rangle,\hspace{-0.4cm}
\label{Zequalexample}
\end{equation}
if $|n_{T_1}\rangle$, $|n_{T_2}\rangle$ and $|n_{A_2}\rangle$ represent the same state in the continuum. 
In the above example we have suppressed the $P$ and $C$ indices for clarity.

\end{itemize}
    
\begin{table}[ht]
\begin{tabular}{ccc|ccc|ccc}
  \hline\hline
  \multicolumn{9}{c}{$\mathbf{p}=(0,0,0)$, $O_h$, $P=\pm$}    \\\hline
  & $\Lambda~(dim)$  &&& \multicolumn{4}{c}{$J$}           &  \\\hline
  & $A_1~(1)$        &&& \multicolumn{4}{c}{$0,~4,...$}    &  \\\hline
  & $T_1~(3)$        &&& \multicolumn{4}{c}{$1,~3,~4,...$} &  \\\hline
  & $T_2~(3)$        &&& \multicolumn{4}{c}{$2,~3,~4,...$} &  \\\hline
  & $E~(2)$          &&& \multicolumn{4}{c}{$2,~4,...$}    &  \\\hline
  & $A_2~(1)$        &&& \multicolumn{4}{c}{$3,~5,...$}           &  \\\hline\hline \\\hline\hline
  \multicolumn{9}{c}{$\mathbf{p}=(0,0,1)$, $Dic_4$}  \\\hline
  & $\Lambda~(dim)$      &&& $|\lambda|^{\tilde \eta}$ &&& $J^P$ (at rest)                  & \\\hline
  &  $A_1~(1)$           &&& $0^+$                     &&& $0^+,~1^-,~2^+,~3^-,~4^+,...$    & \\
  &                      &&& $4$                       &&& $~4^\pm,...$                     & \\\hline
  &  $A_2~(1)$           &&& $0^-$                     &&& $0^-,~1^+,~2^-,~3^+,~4^-,...$    & \\
  &                      &&& $4$                       &&& $~4^\pm,...$                     & \\\hline
  &  $E~(2)$             &&& $1$                       &&& $1^\pm,~2^\pm,~3^\pm,~4^\pm,...$ & \\
  &                      &&& $3$                       &&& $3^\pm,~4^\pm,...$               & \\\hline
  &  $B_1~(1)$           &&& $2$                       &&& $2^\pm,~3^\pm,~4^\pm,...$        & \\\hline
  &  $B_2~(1)$           &&& $2$                       &&& $2^\pm,~3^\pm,~4^\pm,...$        & \\\hline\hline \\\hline\hline
  \multicolumn{9}{c}{$\mathbf{p}=(1,1,0)$, $Dic_2$}  \\\hline
  & $\Lambda~(dim)$      &&& $|\lambda|^{\tilde \eta}$ &&& $J^P$ (at rest)       & \\\hline
  &  $A_1~(1)$           &&& $0^+$                     &&& $0^+,~1^-,~2^+,~3^-,~4^+,...$    & \\
  &                      &&& $2$                       &&& $2^\pm,~3^\pm,~4^\pm,...  $      & \\
  &                      &&& $4$                       &&& $4^\pm,...$                      & \\\hline
  &  $A_2~(1)$           &&& $0^-$                     &&& $0^-,~1^+,~2^-,~3^+,~4^-,...$    & \\
  &                      &&& $2$                       &&& $2^\pm,~3^\pm,~4^\pm,... $           & \\
  &                      &&& $4$                       &&& $4^\pm,... $                     & \\\hline
  &  $B_1~(1)$           &&& $1$                       &&& $1^\pm,~2^\pm,~3^\pm,~4^\pm,...$ & \\
  &                      &&& $3$                       &&& $ 3^\pm,~4^\pm,...$              & \\\hline
  &  $B_2~(1)$           &&& $1$                       &&& $1^\pm,~2^\pm,~3^\pm,~4^\pm,...$ & \\
  &                      &&& $3$                       &&& $ 3^\pm,~4^\pm,...$              & \\\hline\hline
\end{tabular}
\caption{Lattice irreps $\Lambda$ for the symmetry groups corresponding to the momentum 
${\bf p}=(0,0,0),~(0,0,1)$ and~$(1,1,0)$, and continuum quantum numbers that can contribute to 
these lattice irreps. We consider $J\leq 3$ in this work. For the rest frame, the second column 
presents the distribution of spins for $J\leq3$ across different lattice irreps along with the 
next higher spin that can contribute. For ${\bf p}\not =0$, the table lists, which helicities 
($\lambda$) and which $\tilde \eta=P(-1)^J$  can appear in each lattice irrep; $\tilde\eta$ is 
a good quantum number only for $\lambda=0$. In the third column, the $J^P$ with $J\leq4$ that 
can appear in the moving frame irreps are also shown. }\label{tab:2-1}
\end{table}
  
\subsection{Mesons in flight}\label{sec:basicsinflight}

For mesons with non-zero momentum in the infinite volume continuum, the $O(3)$ symmetry group is broken 
to its subgroup, $U(1)$. The eigenstates $|\mathbf{p},J^{PC},\lambda^{\tilde \eta}\rangle$ are labeled 
by the magnitude of helicity $\lambda$ (the projection of the spin component along the direction 
of momentum ${\bf p}/p$) and $\tilde \eta$ parity defined as $\tilde{\eta}=P(-1)^J$. $\tilde \eta$ 
is a good quantum number only for $\lambda=0$ and is related to the reflection in the plane 
containing ${\bf p}$. Unlike in the rest frame, $J^P$ are no longer good quantum numbers and the 
hadron spectrum in moving frames will be a mixture of different $J^P$ for any given $\lambda$ 
\cite{Thomas:2011rh}. Hence, discerning the $J^P$ information of the spectrum in moving frames 
requires more care than for hadrons at rest. 

On the lattice, $\lambda$ and $\tilde \eta$ are no longer good quantum numbers. Lattice eigenstates 
transform according to a given lattice irrep $\Lambda^C$ of the reduced discrete lattice symmetry 
group associated with a particular spatial momentum. We consider ${\bf p}=(0,0,1)$ (in units of 
$\tfrac{2\pi}{N_L}$) with symmetry group $Dic_4$ and ${\bf p}=(1,1,0)$ with symmetry group $Dic_2$. 
Table \ref{tab:2-1} presents all irreps for both cases and lists the values of $\lambda$ and 
$\tilde \eta$~(for $\lambda=0$) that can contribute to each of these lattice irreps. Note that we 
only consider $J\le 3$ ($|\lambda|\leq 3$) in this work, and that the spectrum of every irrep of 
$Dic_2$ and the $E$ irrep of $Dic_4$ receive contributions from two different helicities.

Knowing that a meson with spin $J$ can have helicities $0\leq |\lambda|\leq J$, we list the $J^P$ of
states that can contribute to each irrep in the rightmost column of Table~\ref{tab:2-1}. For example, 
the $A_1$ irrep of $Dic_4$ contains only
$|\lambda|=0$ with an allowed value of $\tilde \eta=+1$. Thus mesons expected to appear in $A_1$
with $|\lambda|=0$ can only have $J^P=0^+,~1^-,~2^+,~3^-$. For $A_1$ in $Dic_2$, we arrive at the 
same pattern of $J^P$ for $|\lambda|=0$, however, states with $|\lambda|=2$ can also contribute. 
In this case $\tilde{\eta}$ is not a good quantum number and hence there is no restriction on the 
parity. States with $J\geq |\lambda|=2$ are possible, leading to the allowed combinations 
$J^P=2^\pm, 3^\pm$.

Clearly, identifying the underlying $J^P$ of energy levels extracted in each lattice irrep is non-trivial 
when studying systems in flight, as a number of states with different $J^P$ can appear in the spectrum 
for each lattice irrep.  To aid this study, we follow the procedure in Ref. \cite{Thomas:2011rh}, which 
is detailed in Section \ref{sec:operators}. Using the continuum interpolators $O^{J^{PC},M}_{i}({\bf p}={\bf 0})$, 
we construct operators at ${\bf p}\ne{\bf 0}$ with good helicity $\lambda$ $(O^{J^{PC},\lambda}_i({\bf p}))$. 
The lattice interpolators respecting the reduced symmetry of the lattice ($O^{[J^{PC},|\lambda|]}_{i,\Lambda^C}$) 
are built from these moving frame continuum interpolators and are therefore expected to have strongest overlap 
with continuum states with quantum numbers $\lambda$ and $\tilde \eta$. Hence our spin identification 
proceeds in two steps : 1) identification of the helicity of lattice energy levels (analogous with the $J$ 
identification for the spectrum at rest), 2) the spin and parity assignments. Below we summarize our `guidelines' 
for spin-parity assignments in a moving frame :

\begin{center}
{\it 1) Identification of $|\lambda|$ and $\tilde \eta$}
\end{center}

\begin{itemize}

\item {\bf Trivial assignments}: Table \ref{tab:2-1} indicates that levels in $A_{1,2}$ of $Dic_4$ 
      can immediately be assigned $\lambda=0$. Similarly levels in $B_{1,2}$ of $Dic_4$ have $|\lambda|=2$.

\item {\bf (Non-)degeneracy of energy levels}: Appearance of near degenerate energy levels (up to cutoff 
and finite volume effects) or absence of a near degenerate partner level across the lattice spectrum 
according to Table \ref{tab:2-1} suggests possible helicity assignments. 

\item {\bf Enhanced $Z$ factors}: Enhanced overlap factors $Z_i^n = \langle O^{[J^{PC},|\lambda|]}_{i,\Lambda^C}|n\rangle$ 
suggest a helicity assignment of $|\lambda|$. Once again we utilize the quantity $\tilde Z_i^n$ as 
defined in Eq. (\ref{Ztilde}). Large overlaps for lattice interpolators subduced from continuum 
interpolators with helicity $|\lambda|$ and small overlaps for lattice interpolators subduced from 
continuum interpolators with helicity $|\lambda'|\ne |\lambda|$ suggest a helicity assignment 
$|\lambda|$ and therefore a minimum continuum spin assignment, {\it i.e.} $J\geq |\lambda|$.

\item {\bf $Z$ factor degeneracy for non-zero helicities}: 
Similar $Z$ factors are expected for a continuum state $|{\bf p};J^{PC},\lambda\rangle$
with interpolators subduced from the same continuum operator $O^{J^{PC},\lambda}_i$ (Eq. (5) of Ref. \cite{Thomas:2011rh}),
\begin{equation}
\langle O^{[J^{PC},|\lambda|]}_{i,\Lambda_1^C}|\mathbf{p},J^{PC},\lambda\rangle \simeq \langle O^{[J^{PC},|\lambda|]}_{i,\Lambda_2^C}|\mathbf{p},J^{PC},\lambda\rangle.\hspace{-0.4cm} 
\label{Zmomequal}
\end{equation}
For example, a $|(1,1,0),3^{PC},3\rangle$ continuum state manifests as two near degenerate energy levels, one each 
in $B_1$ and $B_2$ of $Dic_2$. Both these levels should have degenerate $Z_i^n$ factors for lattice 
interpolators subduced from the same continuum interpolator with $\lambda=3$. Hence 
\begin{equation}
\langle O^{[3^{PC},3]}_{i,B_1^C}|n_{B_1^C}\rangle \simeq \langle O^{[3^{PC},3]}_{i,B_2^C}|n_{B_2^C}\rangle,
\label{Zmomequalexample}
\end{equation}
if $|n_{B_1^C}\rangle$ and $|n_{B_2^C}\rangle$ represent the same $\lambda=3$ state in the continuum.

\end{itemize}

\begin{center}
{\it 2 Identifying $J^P$ from known $|\lambda|^{\tilde \eta}$}
\end{center}

\begin{itemize}

\item {\bf Helicity components}: The spin of the state with a given helicity is constrained to be 
$J\geq |\lambda|$. Different $|\lambda|$ components of a continuum $J^P$ state are constrained to 
have the same energy by Lorentz symmetry \cite{Thomas:2011rh}. These constraints suggest $J^P$ 
assignments by matching the observed pattern of energy levels with the allowed distribution of 
helicity components of a continuum $J^P$ as summarized in Table \ref{tab:2-1}. 

\item {\bf Information in $\tilde Z$ factors:} At rest a continuum interpolator $O^{J^{P},|\lambda|}_{i}({\mathbf p})$
can only overlap with states of spin-parity $J^P$, whereas at non-zero momenta this operator 
can also overlap with states that have other spin and parity \cite{Thomas:2011rh}. If the employed momentum 
and the effects of breaking of the continuum symmetry on the lattice are small, one expects to have dominant 
$\tilde Z$ factors for $O^{[J^{P},|\lambda|]}_{i,\Lambda}({\mathbf p})$ with states that have quantum numbers 
$J^P$ in the continuum at rest. We utilize this expectation not only to confirm the initial spin-parity 
assignments, but also to identify the quantum numbers for remaining ambiguous levels. 

\end{itemize}

\section{Interpolators}\label{sec:operators}
  
Only quark-antiquark ($c\bar c$) interpolators are considered in this work. To construct interpolators 
with good overlaps to physical states, we follow the methods of Refs.~\cite{Dudek:2010wm} 
and~\cite{Thomas:2011rh} for the rest frame and moving frames, respectively. The approach involves first
building interpolators in the continuum with angular momentum, $J$, its $z$-component $M$ and parity $P$ 
at ${\bf p}={\bf 0}$. Using these rest frame interpolators, operators at ${\bf p}\ne{\bf 0}$ and helicity 
$\lambda$ are constructed. These continuum interpolators are then subduced onto their lattice 
counterparts using the respective subduction coefficients. 
If the effects of a finite lattice spacing are small, these subduced lattice operators will have a good overlap 
with the states in the continuum. For completeness, we outline the main steps below and refer the reader 
to Refs.~\cite{Dudek:2010wm,Thomas:2011rh} for further details.

The starting point is a basis of continuum quark bilinear interpolators with spin structures built 
from gamma matrices and single and double covariant derivatives,
\begin{equation}\label{eq:bilinears}
\bar c(x) \Gamma c(x), \quad \bar c(x) \Gamma \overleftrightarrow{D}_j c(x), \quad \bar c(x) \Gamma \overleftrightarrow{D}_j \overleftrightarrow{D}_k c(x).
\end{equation}
The derivatives $\overleftrightarrow{D}_j = \overrightarrow{D}_j-\overleftarrow{D}_j$ and the vector 
combinations of gamma matrices are formed into circular bases \cite{Thomas:2011rh}. A combination of 
these bilinears with O(3) Clebsch-Gordan coefficients~($C^{CG}_i$) leads to continuum operators $O^{J^{PC},M}_i({\bf p})$, 
which have definite $J$, $M$ and $P$ for ${\bf p}=0$. The subscript $i$ is an index identifying 
the continuum interpolator in the rest frame. For example, utilizing the two-derivative bilinears in 
Eq.~(\ref{eq:bilinears}) one can form operators with
$J\le 3$:
\begin{align}\label{eq:3-1}
O^{J^{PC},M}_i({\bf p})= &\sum_{m_1,m_2,m_3} C^{CG}_i(m_1,m_2,m_3;M)\times\nonumber\\
&\sum_{{\bf x}} \bar
c(x) \Gamma_{m_1} \overleftrightarrow{D}_{m_2}\overleftrightarrow{D}_{m_3}c(x) e^{i{\bf p\cdot x}}.
\end{align}
Similarly using one-derivative bilinears in Eq.~(\ref{eq:bilinears}), one can construct $J\le 2$ operators
and $J\le 1$ interpolators can be built using the local bilinears. 

At ${\bf p=0}$ the projection onto lattice interpolators for each lattice representation $\Lambda$ 
proceeds via subduction coefficients $S^{J,M}_{\Lambda,\mu}$ of the $O_h$ group
\begin{equation}\label{eq:3-2}
O^{[J^{PC}]}_{i,\Lambda^C,\mu}({\bf p=0})=\sum_M S^{J,M}_{\Lambda,\mu} ~O^{J^{PC},M}_i({\bf p=0}),
\end{equation}
where $\mu$ indicates the row of the lattice irrep and operators on the r.h.s are understood to be 
discrete versions of the continuum ones.  See Appendix A of Ref.~\cite{Dudek:2010wm} for subduction 
coefficients for $J\le 4$.

At finite ${\bf p}$ one first forms continuum operators of definite helicity $\lambda$: 
\begin{equation}\label{eq:3-3}
O^{J^{PC},\lambda}_i({\bf p})= \sum_M {\cal D}^{(J)*}_{M,\lambda}(R)\;O^{J^{PC},M}_i({\bf p}),
\end{equation}
where ${\cal D}^{(J)*}_{M,\lambda}(R)$ is the Wigner-${\cal D}$ matrix and R refers to the rotation 
from $(0,0,|{\bf p}|)$ to ${\bf p}$. For the trivial case of ${\bf p}=(0,0,|{\bf p}|)$,
${\cal D}^{(J)*}_{M,\lambda}(R)$ is a unit matrix and $O^{J^{PC},M}_i({\bf p}) = O^{J^{PC},\lambda}_i({\bf p})$. 
As discussed in Ref. \cite{Thomas:2011rh}, in order to ensure consistency between different momentum 
directions, $R$ is split into two parts $R=R_{\rm lat}R_{\rm ref}$: a reference rotation $R_{\rm ref}$ 
that takes $(0,0,|{\bf p}|)$ to ${\bf p}_{\rm ref}$ and a lattice rotation $R_{\rm lat}$ from 
${\bf p}_{\rm ref}$ to ${\bf p}$. We take ${\bf p}_{\rm ref}=(0,0,1)$ and $(1,1,0)$ for $Dic_4$ and 
$Dic_2$, respectively. Finally, these continuum operators are subduced to obtain the lattice interpolators
\begin{equation}\label{eq:3-4}
O^{[J^{PC},|\lambda|]}_{i,\Lambda^C,\mu}({\bf p})=  \sum_{\hat{\lambda}=\pm|\lambda|} S_{\Lambda,\mu}^{\tilde{\eta},\hat{\lambda}}~O^{J^{PC},\hat{\lambda}}_i({\bf p}).
\end{equation} 
The subduction coefficients, $S_{\Lambda,\mu}^{\tilde{\eta},\hat{\lambda}}$ for $Dic_2$ and $Dic_4$
can be found in Table II of Ref.~\cite{Thomas:2011rh}. 
We construct the interpolators in all momentum polarizations. The correlation functions of different 
momentum polarizations for a given $|{\bf p}|$ are averaged before the GEVP analysis (discussed in 
section \ref{sec:setup}).

To illustrate the above procedure, consider the example of an interpolator in the (one-dimensional) $B_2$ irrep 
of $Dic_4$ ({\it i.e.} ${\bf p}=(0,0,1)$), which receives contributions from $\lambda=\pm2$~(see 
Table~\ref{tab:2-1}). We start with a continuum interpolator in the rest frame, 
$O^{2^{++},M}_i({\bf p}=0)$. For the choice of momentum ${\bf p}=(0,0,1)$, the Wigner matrix in 
Eq.~(\ref{eq:3-3}) is a unit matrix. Hence the continuum interpolator in the moving frame is 
simply $O^{2^{++},\lambda}_i({\bf p})$. The subduction coefficients for $B_2$ of $Dic_4$ read~\cite{Thomas:2011rh},
$$ S_{B_2}^{\tilde{\eta},\hat{\lambda}}=(\delta_{s,+}-\tilde{\eta}\delta_{s,-})/\sqrt{2}= \pm 1/\sqrt{2} ~~\ \mathrm{for}\  \hat{\lambda}=\pm 2, \nonumber $$
with $\tilde{\eta}=P(-1)^J=+1$ in our example and $s\equiv\text{sign}(\lambda)$. 
This leads to the lattice interpolator
\begin{equation}
O^{[2^{++},2]}_{i,B_2^+}({\bf p}) = (O^{2^{++},+2}_i({\bf p})+O^{2^{++},-2}_i({\bf p}))/\sqrt{2}.
\end{equation}


\section{Lattice setup}\label{sec:setup}

The charmonium spectrum has been determined on an $N_f = 2+1$ ensemble produced by the CLS 
consortium \cite{Bruno:2014jqa,Mohler:2017wnb}, labeled as U101. The configurations have 
been generated with a non-perturbatively $\mathcal{O}(a)$ improved Wilson fermion action 
and a Symanzik gauge action. The pion and the kaon masses are $m_\pi \simeq  280$ MeV and 
$m_K \simeq  467$ MeV respectively. The U101 ensemble lies along the 
$\textrm{Tr}(m)=2m_{u/d}+m_s =\textrm{const}$ line, meaning that the strange quark mass 
becomes heavier and the light quark mass lighter as the physical point is approached. The 
volume of the lattice is $24^3\times 128$ and the lattice  spacing is $a=0.08636(98)(40)$ fm 
\cite{Bruno:2016plf} giving 
a physical spatial volume of $(2.07~ \textrm{fm})^3$. The gauge and fermion fields fulfill 
open boundary conditions in the time direction and translational invariance is only expected 
in the bulk \cite{Luscher:2011kk}. Hence we measure two-point correlation functions in 
the middle of our lattice at least 28 time slices away from the boundaries. No effects related 
to the finite temporal extent are seen in this time interval for the pion and charmed meson 
correlation functions. We utilize a total of 1638 source time slices from 255 configurations 
to compute the correlation functions.

The charm quark is quenched in our simulations and vacuum charm quark loops are absent in the 
theory. The charm quark mass is therefore set independently when measurements are performed. 
There are various possible observables that can be used for tuning the charm quark mass. Our 
long-term objective is the determination of the properties of exotic charmonium resonances, 
therefore the position of the $\bar{D}$-$D$ decay threshold needs to be taken into account. In 
particular, a pion mass heavier than its experimental value might result in a situation where 
a physical resonance will appear as a bound state. Another important issue is how the 
properties of the exotic charmonium states depend on the charm quark mass. We have therefore 
determined the charmonium spectrum for two different values of the charm quark mass, corresponding 
to $\kappa_c = 0.12315$ and $\kappa_c=0.12522$, resulting in $D$-meson masses approximately 80 MeV 
above and below the physical value, respectively. In the present work, we focus on the results 
from $\kappa_c=0.12522$, and we will utilize the data from $\kappa_c = 0.12315$ for our future 
meson scattering analysis.

We employ the distillation method~\cite{Peardon:2009gh} to compute matrices of correlation functions 
between a basis of meson interpolating fields. This method is powerful enough to implement local 
and non-local hadron interpolators as well as multi-hadron interpolators, and to perform 
definite non-zero momentum projections at the source and at the sink, which are crucial in a study 
of scattering amplitudes and resonances. This is achieved with a reasonable amount of computational 
resources and 
disk space requirements. Distillation is equivalent to standard Gaussian quark field smearing 
algorithms, written in terms of the eigenvectors of the gauge covariant lattice Laplacian in three 
dimensions. We construct our fermion sources from 90 eigenvectors and we employ \emph{full} distillation, 
meaning that we compute the quark propagation between the full set of eigenvectors at the source and 
at the sink.

The computation of a given entry of the correlation matrix proceeds in four steps. First, the Arnoldi 
algorithm provides the smallest eigenvectors and eigenvalues of the Laplacian operator on all time slices 
$t$ fulfilling the equation
\begin{equation}
 \nabla^2_{[t]} \Psi_k(t) = \lambda^k_t \Psi_k(t),
\end{equation}
where spatial and color indices are suppressed and $k$ labels the eigenvector index.
Second, we solve the Dirac equation
\begin{equation}
 D(t,t',\alpha,\alpha';m) \eta_k(t',\alpha') = \psi_k(t, \alpha)\,
\end{equation}
to obtain the quark propagators $\eta^k(t',\alpha')$ from each source $\psi_k(t, \alpha)$. The source is 
zero everywhere except for time slice $t$ and spin component $\alpha$, where it is equal to the eigenvector 
$\Psi_k(t)$. The perambulators $\tau(t,t',\alpha,\alpha',k,k')$ are constructed from the $\eta_k(t',\alpha')$ using 
\begin{equation}
 \tau(t,t',\alpha,\alpha',k,k') = \psi_{k}(t, \alpha)^\dag \eta_{k'}(t',\alpha').
\end{equation}

In the third step, we compute the matrices $\phi$ 
\begin{equation}
 \phi(t,\alpha,\alpha',k,k') = \psi_{k'}(t, \alpha)^\dag O_{\alpha\alpha'} \psi_{k}(t, \alpha')\,.
\end{equation}
The operator $O_{\alpha\alpha'}(k,k')$ has a spin structure and possibly also
covariant derivatives acting on the Laplacian eigenvectors, therefore in 
general $\phi$ is not diagonal in the distillation space. The matrices $\phi$
carry the information about the quantum numbers of the operators that define the correlation function.  

Both $\phi$ and $\tau$ can be considered as squared matrices of dimension $(90\times4)\times(90\times4)$ defined on each time slice or connecting two time slices.
From these matrices, we can finally compute the sum of all the Wick contractions, neglecting only the charm annihilation diagrams that are OZI suppressed. For instance, a simple $\bar{q}q$-meson correlation function reads explicitly
\begin{equation}
 C(t,t') = \textrm{Tr}\left\{\phi(t) \tau(t,t')\phi(t')^\dag \tau(t',t)^\dag\right\}\,.
\end{equation}

The correlation matrices (Eq. (\ref{eq:2-1})) are computed for all irreps in Table \ref{tab:2-1} using
$\bar cc$ operators constructed as in Section \ref{sec:operators}. In Table \ref{tab:5-1},
we list the number of interpolators employed in each of the irreps in the three inertial frames we study.
For multi-dimensional irreps, we restrict the computation of correlation matrices to one single row
of the irrep. For brevity, we drop the subscript $\mu$ from the lattice operators used in 
Eqs. (\ref{eq:3-2}) and (\ref{eq:3-4}) throughout this section. Energies $E_n$ and overlaps 
$Z_i^n=\langle O_{i,\Lambda^C}|n\rangle$ of the lattice levels are extracted from solutions of 
the GEVP
\begin{equation}
C(t)u^n(t)=\lambda^n(t)C(t_0)u^n(t) 
\end{equation}
with $t_0=2$. $E_n$ are extracted by one or two exponential correlated fits to $\lambda^n(t)$, 
whereas $Z_i^n$s are extracted from constant fits to a plateau in the function,
\begin{equation}
Z_i^n(t) = e^{E_n t/2}C_{ij}(t)u_j^n(t)/|C(t)^{1/2}u^n(t)|.
\end{equation}
The quality of the fits to $\lambda^n(t)$ is illustrated for the example of the $E^-$ irrep of $Dic_4$ in 
Figure~\ref{fig:app-2} of Appendix~\ref{app:fits}. We illustrate the plateaus in $Z_i^n$ for selected 
interpolators and the respective fits for the example of $n=1$, 2 and 6 levels in the spectrum for 
the $E^-$ irrep of $Dic_4$ in Figure~\ref{fig:app-3} of Appendix~\ref{app:fits}.

\begin{table}[ht]
\begin{tabular}{cccccc|ccccccc|cccccc|cccccc}
  \hline
  \multicolumn{25}{c}{$\mathbf{p}=0$, $O_h$, $\Lambda^{PC}$}  \\\hline
  & $A_1^{++}$ &&&  5 &&& $A_1^{+-}$ &&&  3 &&&& $A_1^{-+}$ &&&  7 &&& $A_1^{--}$ &&&  2 & \\
  & $T_1^{++}$ &&&  8 &&& $T_1^{+-}$ &&&  8 &&&& $T_1^{-+}$ &&&  4 &&& $T_1^{--}$ &&& 12 & \\
  & $T_2^{++}$ &&&  6 &&& $T_2^{+-}$ &&&  4 &&&& $T_2^{-+}$ &&&  5 &&& $T_2^{--}$ &&&  5 & \\
  &   $E^{++}$ &&&  5 &&&   $E^{+-}$ &&&  3 &&&&   $E^{-+}$ &&&  5 &&&   $E^{--}$ &&&  3 & \\
  & $A_2^{++}$ &&&  1 &&& $A_2^{+-}$ &&&  0 &&&& $A_2^{-+}$ &&&  1 &&& $A_2^{--}$ &&&  2 & \\\hline\hline
\multicolumn{12}{c}{$\mathbf{p}=(0,0,1)$, $Dic_4$, $\Lambda^C$}&&\multicolumn{12}{c}{$\mathbf{p}=(1,1,0)$, $Dic_2$, $\Lambda^C$}\\\hline
  & $A_1^+$    &&& 14 &&& $A_1^-$    &&& 18 &&&&  $A_1^+$    &&& 25 &&& $A_1^-$   &&& 27 & \\
  & $A_2^+$    &&& 20 &&& $A_2^-$    &&& 12 &&&&  $A_2^+$    &&& 31 &&& $A_2^-$   &&& 21 & \\
  & $B_1^+$    &&& 11 &&& $B_1^-$    &&&  9 &&&&  $B_1^+$    &&& 23 &&& $B_1^-$   &&& 29 & \\
  & $B_2^+$    &&& 11 &&& $B_2^-$    &&&  9 &&&&  $B_2^+$    &&& 23 &&& $B_2^-$   &&& 29 & \\
  & $E^+$      &&& 23 &&& $A^-$      &&& 29 &&&&             &&&    &&&           &&&    & \\\hline
\end{tabular}
\caption{Number of interpolators with up to two derivatives used in computing correlation matrices 
of each lattice irrep in the rest frame (top) and in the moving frames (bottom) with momentum 
${\bf p}=(0,0,1)$ on the left and ${\bf p}=(1,1,0)$ on the right. }\label{tab:5-1}
\end{table}

\section{Results}\label{sec:results}

In this section, we present the charmonium spectra for all lattice irreps in the three 
inertial frames considered. Using this energy spectrum and the corresponding operator state 
overlaps, we illustrate the spin identification procedure outlined in Section \ref{sec:idproc}
and present the spin-identified charmonium spectrum in the three frames studied. 
  
\begin{figure*}[tb]
\begin{center}
\includegraphics[height=0.7cm,width=10cm]{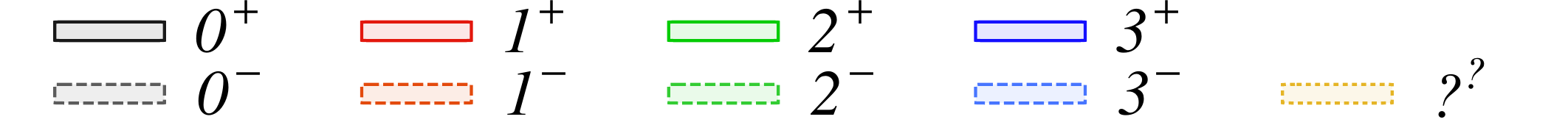}\\
$\quad$\\
\includegraphics[height=5.5cm,width=8.9cm]{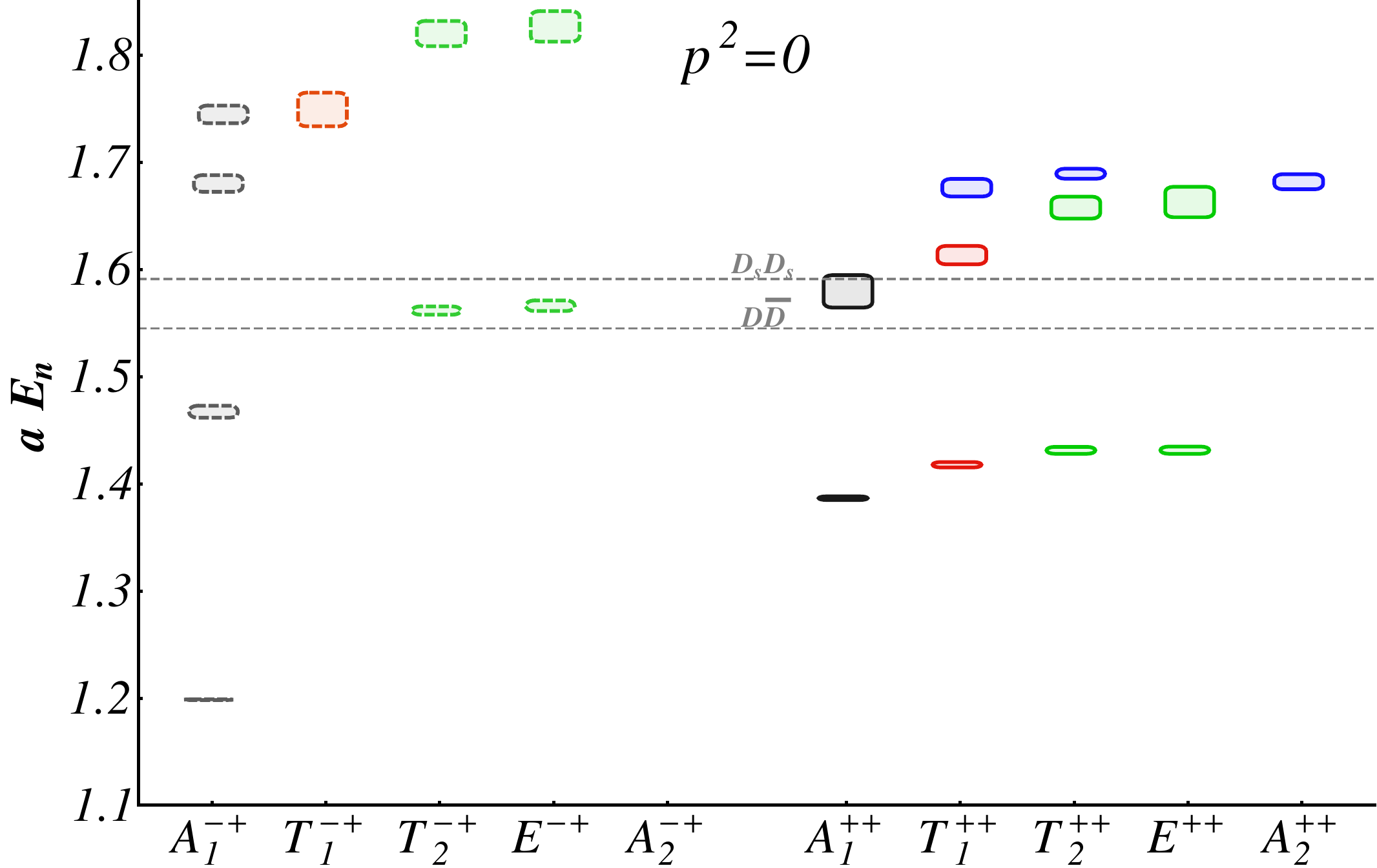}
\includegraphics[height=5.5cm,width=8.9cm]{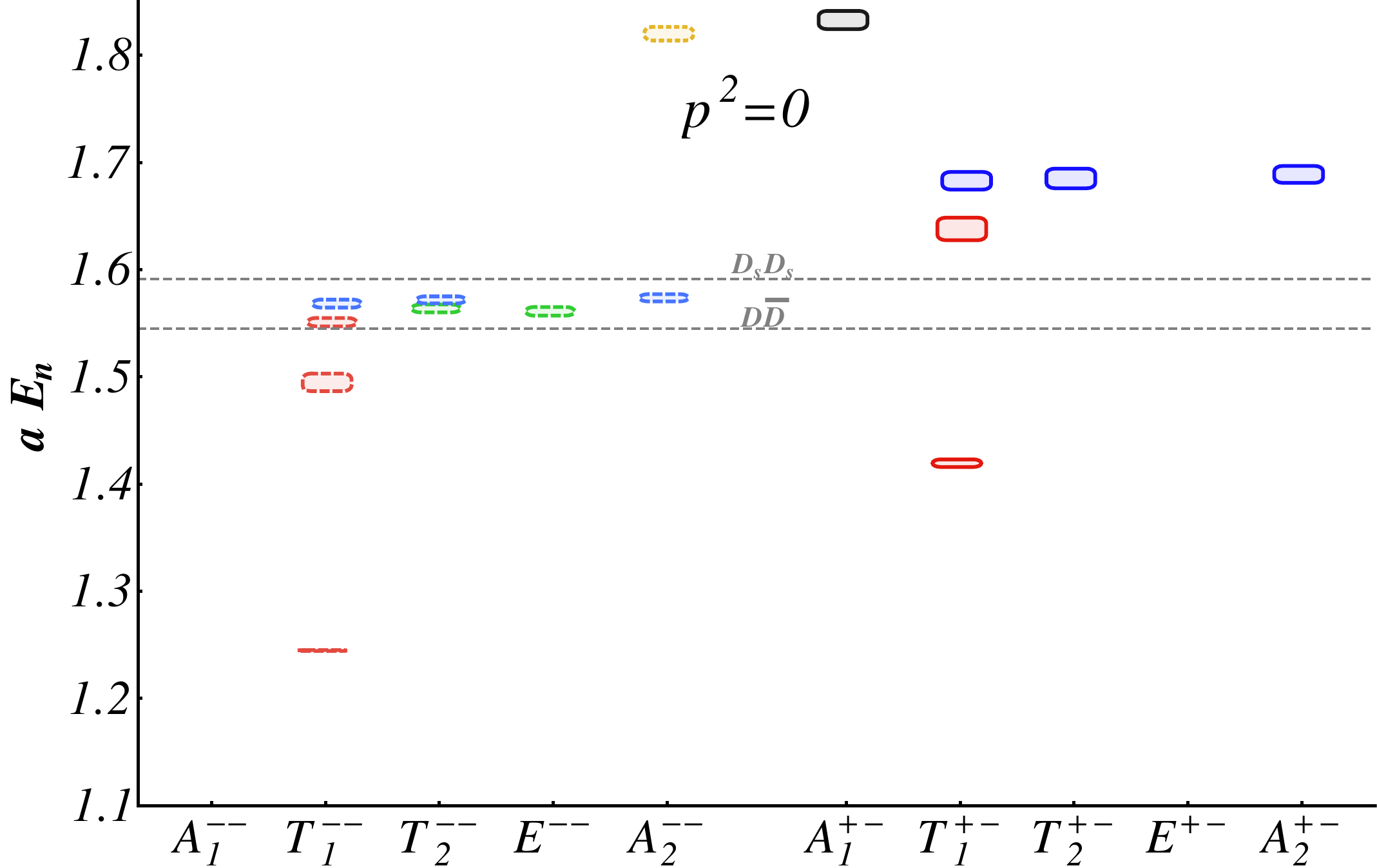}
\end{center}
\caption{ The charmonium spectrum with assignments of $J^P$ in the rest frame for different lattice irreps, $\Lambda^{PC}$.
Different colors indicate the $J^P$ of states according to the color-coding in (\ref{eq:colors}). }\label{fig:5-1}
\end{figure*}

\subsection{Charmonia at rest}\label{sec:results-rest}

In Fig. \ref{fig:5-1}, we present the spin-identified charmonium spectrum in the rest frame for all 
lattice irreps, $\Lambda^{PC}$. The color coding for the spectrum is similar to that used in Ref. 
\cite{Thomas:2011rh}, allowing for an immediate comparison of the qualitative features : 
 \begin{align}\label{eq:colors}
    &J=0:&   &\mathrm{solid\ black}\  P\!=\!+\            &&\mathrm{dashed\ grey}\ P\!=\!- \\
    &J=1:&   &\mathrm{solid\ red}\  P\!=\!+\      &&\mathrm{dashed\ red}\ P\!=\!-\nonumber\\
    &J=2:&   &\mathrm{solid\ green}\  P\!=\!+\    &&\mathrm{dashed\ green}\ P\!=\!-\nonumber\\
    &J=3:&   &\mathrm{solid\ blue}\  P\!=\!+\     &&\mathrm{dashed\ blue}\ P\!=\!-\nonumber\\
    &J^P &   &\mathrm{unidentified}: \               &&\mathrm{dotted\ orange}\nonumber
 \end{align}
Bright colors and solid outlines are used to highlight positive parity states, whereas light colors 
with dashed outlines are used for negative parity states. Lattice levels with ambiguous signatures 
are indicated using orange boxes with dotted outlines. 

The spin assignment of the charmonium spectrum in the rest frame is relatively straightforward and follows 
immediately from Table \ref{tab:2-1} and from the guidelines in Section \ref{sec:basicsrest}. 

{\bf Trivial assignments } : Among $J\leq 3$, the spectrum in the $A_1$ irreps can contain only $J=0$ 
states, hence $J=0$ assignment for these levels is immediate. Any effects of higher spins in these 
levels can only be due to $J\ge4$ states, which we assume to be negligible for two reasons: Firstly, 
no such low lying high spin charmonium states have been discovered. Secondly, we have not employed 
interpolators with such continuum quantum numbers. Analogously, levels in the spectrum of the $E$ 
and $A_2$ irreps can only be assigned $J=2$ and $J=3$, respectively.

{\bf Energy degeneracies }: Continuum states with $J=1$ can appear only in $T_1$. Hence initial $J=1$ 
assignments can be made for the lowest two levels in $T_1$ irreps (one in $T_1^{-+}$, which is an 
exotic quantum number), as they appear alone with no equivalent partner levels across other lattice 
irreps with the same $P$ and $C$. The next higher spin that can occur in $T_1$ is $J=3$, with near 
degenerate energy levels appearing in the $T_2$ and $A_2$ spectra. Thus among the third and the fourth 
levels in $T_1^{--}$ ($aE_n\sim 1.57$ in Fig. \ref{fig:5-1}), the fourth level can be assigned $J=3$, 
as is supported by the near degenerate energy levels in $T_2^{--}$ and $A_2^{--}$. $J=1$ is assigned to 
the third level with no remaining near degenerate partner levels in $T_2^{--}$ and $A_2^{--}$.
In general, the degeneracy of energy levels representing the same continuum state provide a handle on 
initial spin assignments of states with $J\ge2$.

\begin{figure}[tb]
\includegraphics[width=0.45\textwidth,clip]{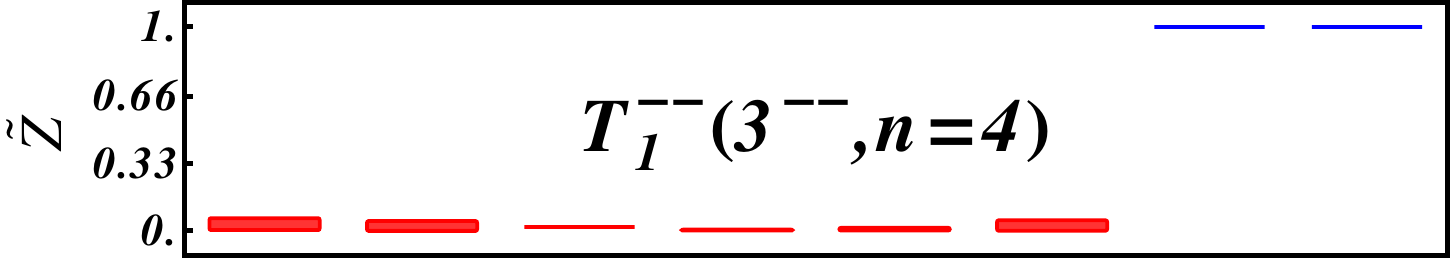}
\includegraphics[width=0.45\textwidth,clip]{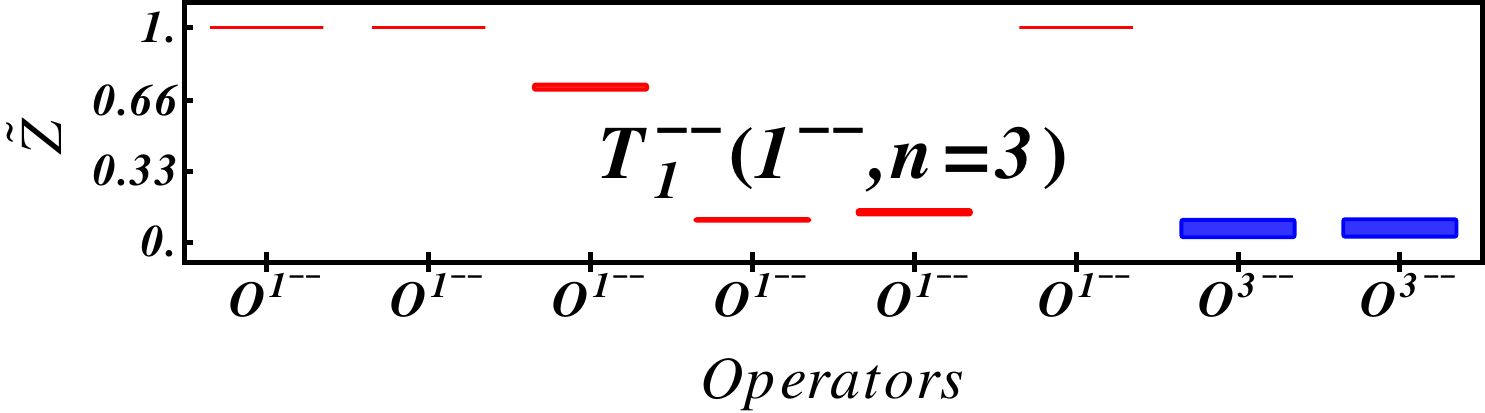}
\caption{$\tilde Z_i^n$ factors for a subset of operators from the $T_1^{--}$ irrep to
demonstrate the spin assignment of the third and fourth states.}\label{fig:5-2}
\end{figure}

\begin{figure}
\includegraphics[width=0.45\textwidth,clip]{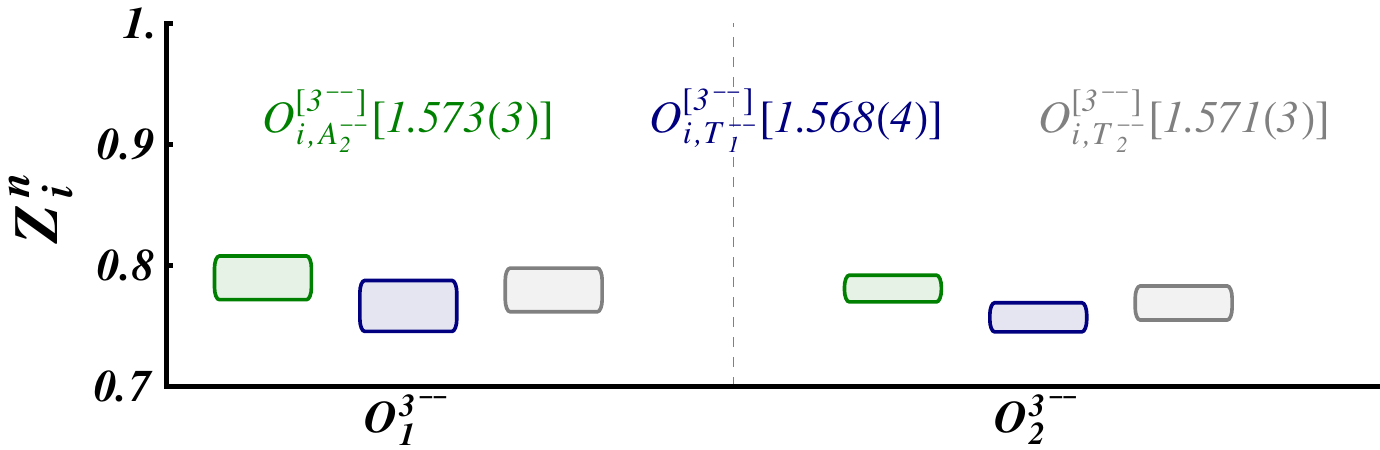}
\includegraphics[width=0.45\textwidth,clip]{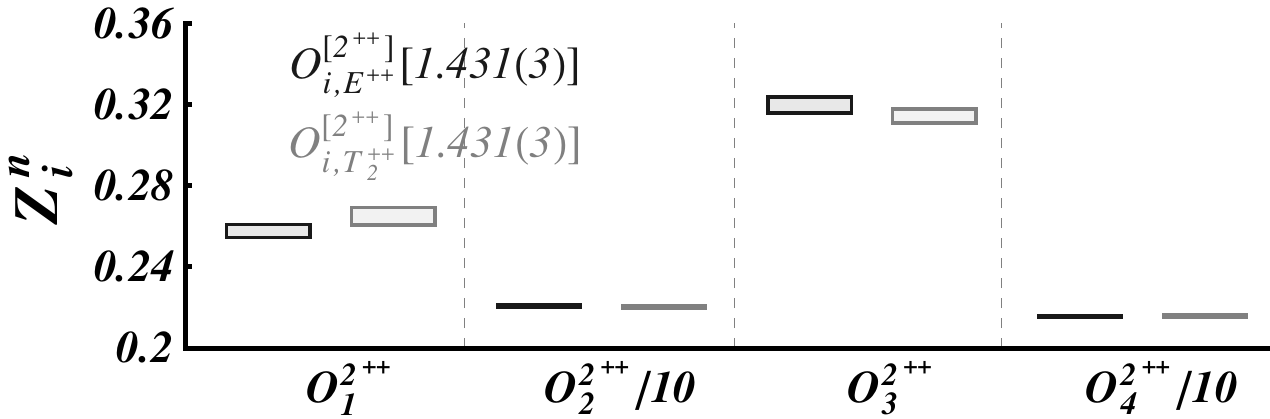}
\caption{Comparison of the $Z_i^n$ factors used to confirm the spin assignments for $J\ge 2$. The upper figure 
shows the comparison for the state identified as $J^{PC} = 3^{--}$ and the lower figure presents a similar 
study for the state identified as $\chi_{c2}(1P)$. The color coding for the lattice irreps are indicated in 
each figure, where the numbers given within square brackets are the respective energies. }\label{fig:5-3}
\end{figure}

\begin{figure*}[bht]
\begin{center}
\includegraphics[height=6cm,width=6cm]{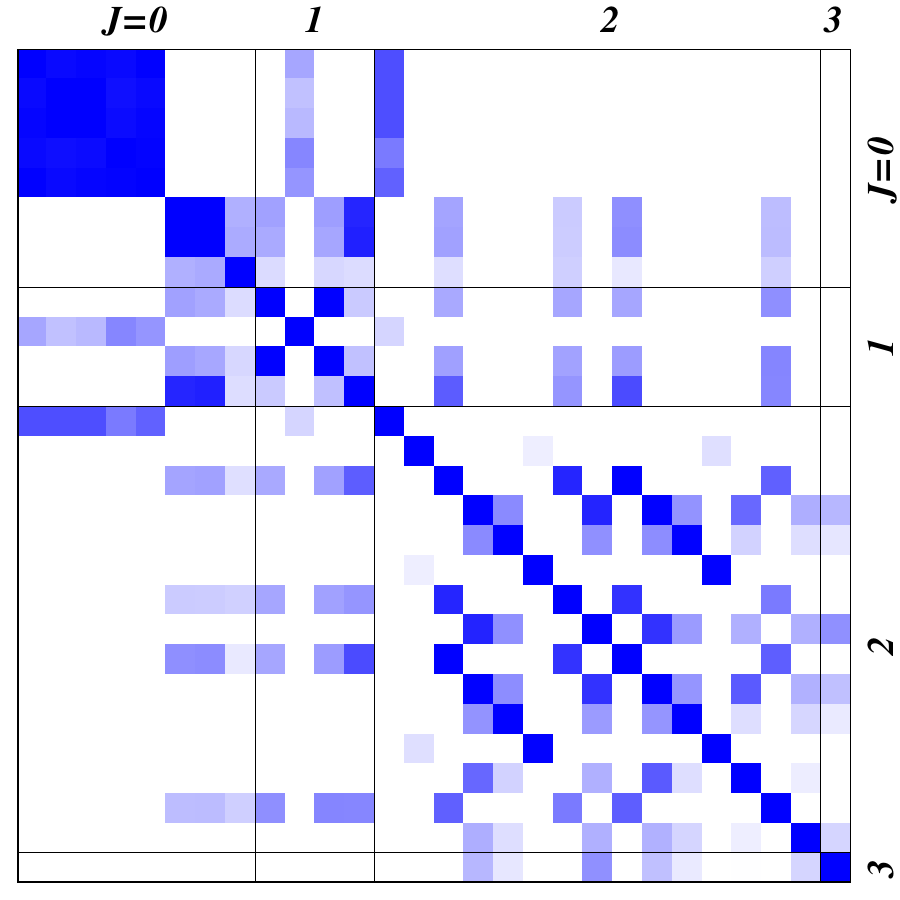}\qquad
\includegraphics[height=3cm,width=0.9cm]{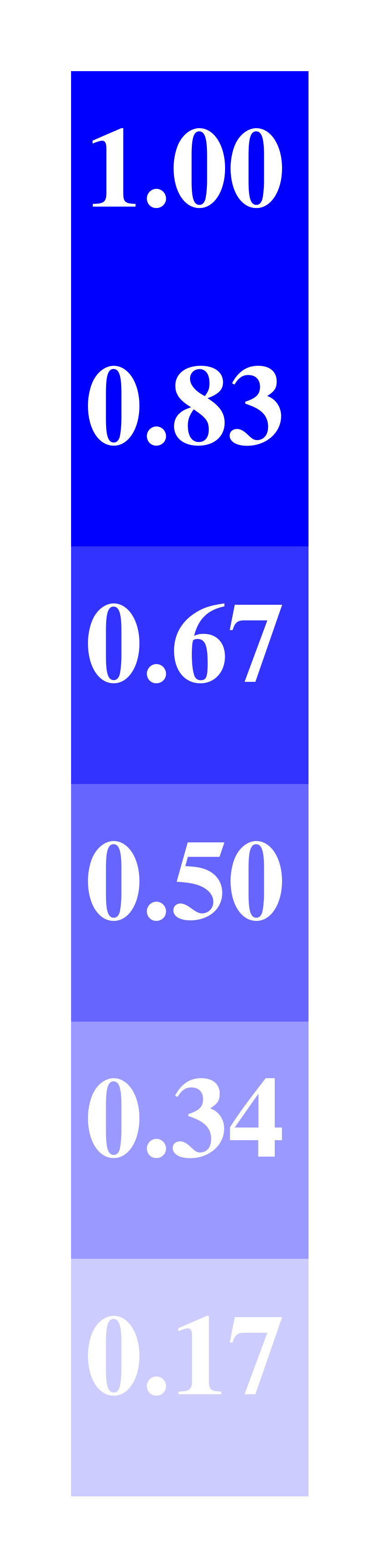} \qquad
\includegraphics[height=6cm,width=6cm]{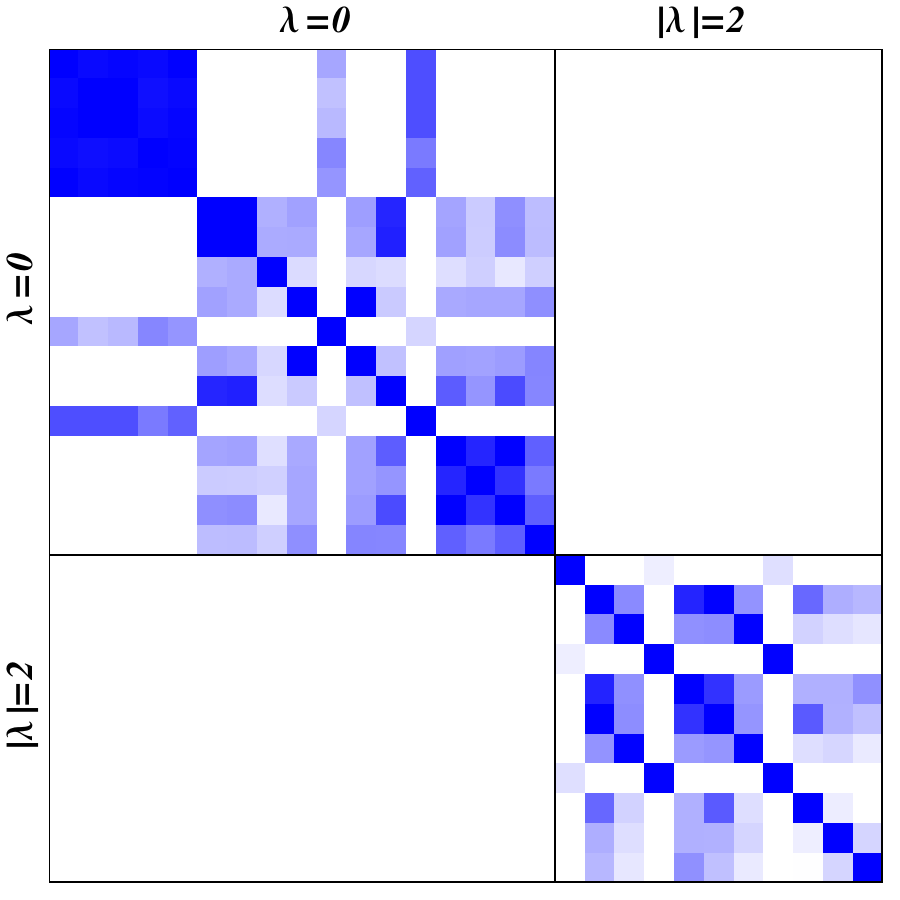}$\qquad$
\end{center} 
\caption{Normalized correlation matrices ($C_{ij}/\sqrt{C_{ii}C_{jj}}$) of the $A_1^{+}$ irrep in $Dic_2$
on time slice 5. Left : operators ordered by spin $J=0,1,2,3$ of the original continuum rest frame 
interpolator. Right : operators $O^{[J,P,|\lambda|]}$ are ordered by helicity $|\lambda|=0,2$.
Vertical and horizontal black lines separate operators of various $\lambda$ or $J$. }\label{fig:5-4}  
\end{figure*}

{\bf $Z$ factor enhancements }: Lattice levels resembling $J\ge2$ often appear in dense energy bands 
along with states with different spins. In such cases, disentangling the quantum number information 
solely from the energies can be delicate. Initial spin assignments need to be further confirmed by 
investigating the $\tilde Z$ factors (Eq. (\ref{Ztilde})). In Fig. \ref{fig:5-2}, we show the 
$\tilde Z$ factors for a subset of operators from the $T_1^{--}$ irrep to demonstrate the spin assignment for the third 
and fourth levels. The $x$-axis labels refer to the continuum interpolators $O^{J^{PC}}$, from 
which the respective lattice interpolators are built. Other indices are suppressed for brevity. 
Clearly, the third level ($n=3$) has the largest and dominant $\tilde Z$ factors for lattice interpolators 
subduced from $J=1$ continuum interpolators, so this state is assigned spin $J=1$. The fourth state 
($n=4$) has dominant $\tilde Z$ factors for lattice interpolators subduced from $J=3$ and  is assigned spin 
$J=3$. 

 {\bf $Z$ factor degeneracies }: By studying the expected degeneracies in $Z$ factors (Eq. (\ref{Zequal})),
one can further confirm the $J\geq 2$ assignments. In Fig. \ref{fig:5-3}, we show such a comparative 
study of $Z$ factors for the lattice levels identified as $3^{--}$ (upper) and $2^{++}$ (lower). The 
upper plot confirms the expected degeneracies for a $J=3$ state given in Eq. (\ref{Zequalexample}) for 
two different continuum interpolators (indicated by $O^{3^{--}}_1$ and $O^{3^{--}}_2$ in the figure). Hence, a $J=3$ assignment for 
the three levels in $T_1^{--}$, $T_2^{--}$ and $A_2^{--}$ at $aE_n\sim1.57$ is immediate. Similarly, 
the lower plot confirms the degeneracy in $Z$ factors $$\langle O^{[2^{++}]}_{i,E^+}|n_{E^+}\rangle\simeq  \langle O^{[2^{++}]}_{i,T_2^+}|n_{T_2^+}\rangle$$
for four different choices of initial continuum operator $O^{2^{++}}_i$, each of which is subduced to 
$T_2^{++}$ and $E^{++}$. This confirms a $J=2$ assignment for the levels at $aE_n\sim1.431$ in 
these two irreps ({\it c.f.} Fig. \ref{fig:5-1}).

By considering the pattern of energies and operator state overlaps of the lattice spectrum in the rest frame,
we arrive at the spin-identified charmonium spectrum as depicted in Fig. \ref{fig:5-1}. Within
the single hadron approach, we assume the lattice energy levels represent the single hadron spectrum
neglecting any effects of allowed strong decay thresholds. The interpretation of energy levels
close to scattering thresholds, such as the first excitation in the $T_1^{++}$ spectrum, are subtle. 
The presence of the nearby non-interacting levels   $D\bar D^*$ might have a strong influence on 
the determination of the energy of these excitations. We postpone these discussions to a follow 
up publication.

\subsection{Charmonia in flight}\label{sec:results-flight}

$|\lambda|-$assignments of the lattice levels can be made following the guidelines discussed 
in Section \ref{sec:basicsinflight}. This is straightforward if the different helicity sectors are decoupled, 
{\it i.e.} if the breaking of $U(1)$ symmetry is small. As the lattice interpolators for mesons in flight 
are subduced from continuum interpolators with good helicity, this is indeed observed in our correlation
matrices. To demonstrate this, we present  the normalized correlation matrix 
($C_{ij}/\sqrt{C_{ii}C_{jj}}$) on time slice 5 for the $A_1^+$ irrep in $Dic_2$ in Fig. \ref{fig:5-4}. 
A linear color gradient is used from deep blue representing the largest value (unity) to white representing 
the smallest value (zero). On the left, we present the correlation matrix with the operators ordered based on 
the $J$ of the continuum rest frame interpolators used to build the continuum helicity interpolators. 
Strong cross-correlations between different $J$ sectors are evident from the figure.
In the right figure, we order the interpolators such that $\lambda=0$ (operators from 1 to 17) are 
collected before $|\lambda|=2$ (operators from 18 to 28). Clearly the correlation matrix is almost completely block 
diagonal in helicity, indicating a clean signature for the helicity of the lattice levels. Hence in what 
follows, we first discuss the determination of the helicity-identified lattice spectrum. Then, we discuss our 
$J^P$ assignments based on the pattern of energies and overlaps.

\begin{center}
{\it Identification of $|\lambda|$ and $\tilde \eta$}
\end{center}

\begin{figure*}[tb]
\begin{center}
\includegraphics[height=10cm,width=16.2cm]{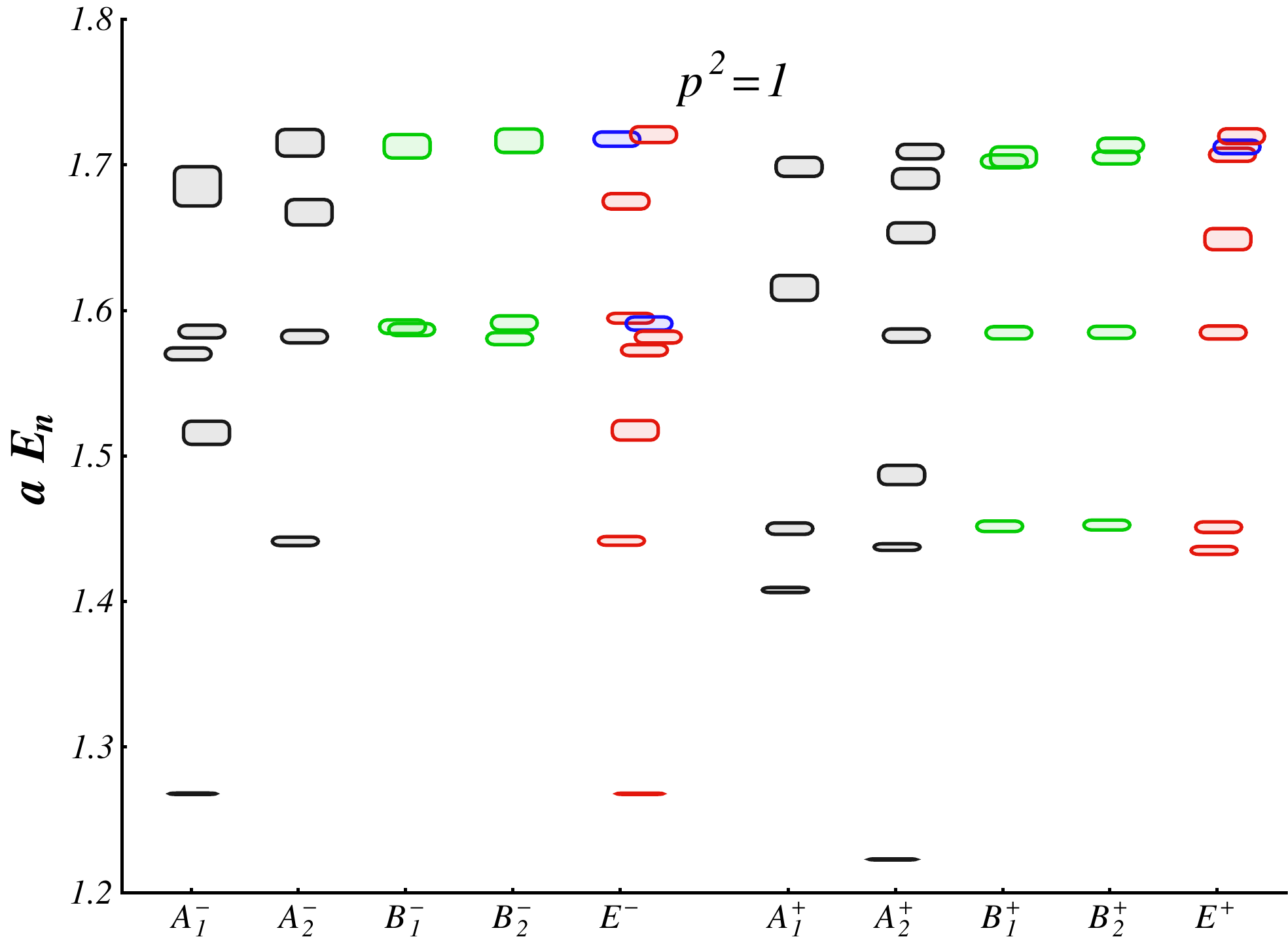}
\includegraphics[scale=0.4]{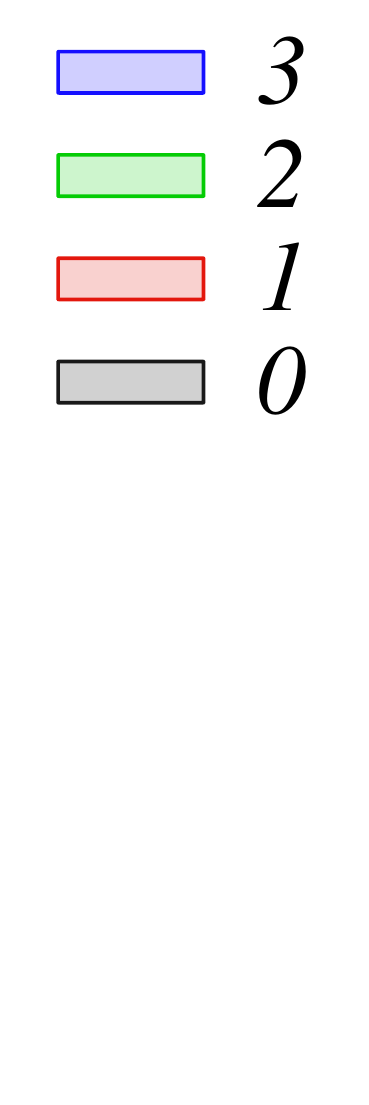}
\end{center}
\caption{$|\lambda|$-identified charmonium spectrum in the moving frame with $\mathbf{p}=(0,0,1)$. 
All irreps $\Lambda^{C}$ of the corresponding little group $Dic_4$ are presented (Table \ref{tab:2-1}). 
The colors indicate $|\lambda|$ of states according to the color-coding (\ref{eq:colorsh}). }\label{fig:5-5}
\begin{center}
\includegraphics[height=10cm,width=16.2cm]{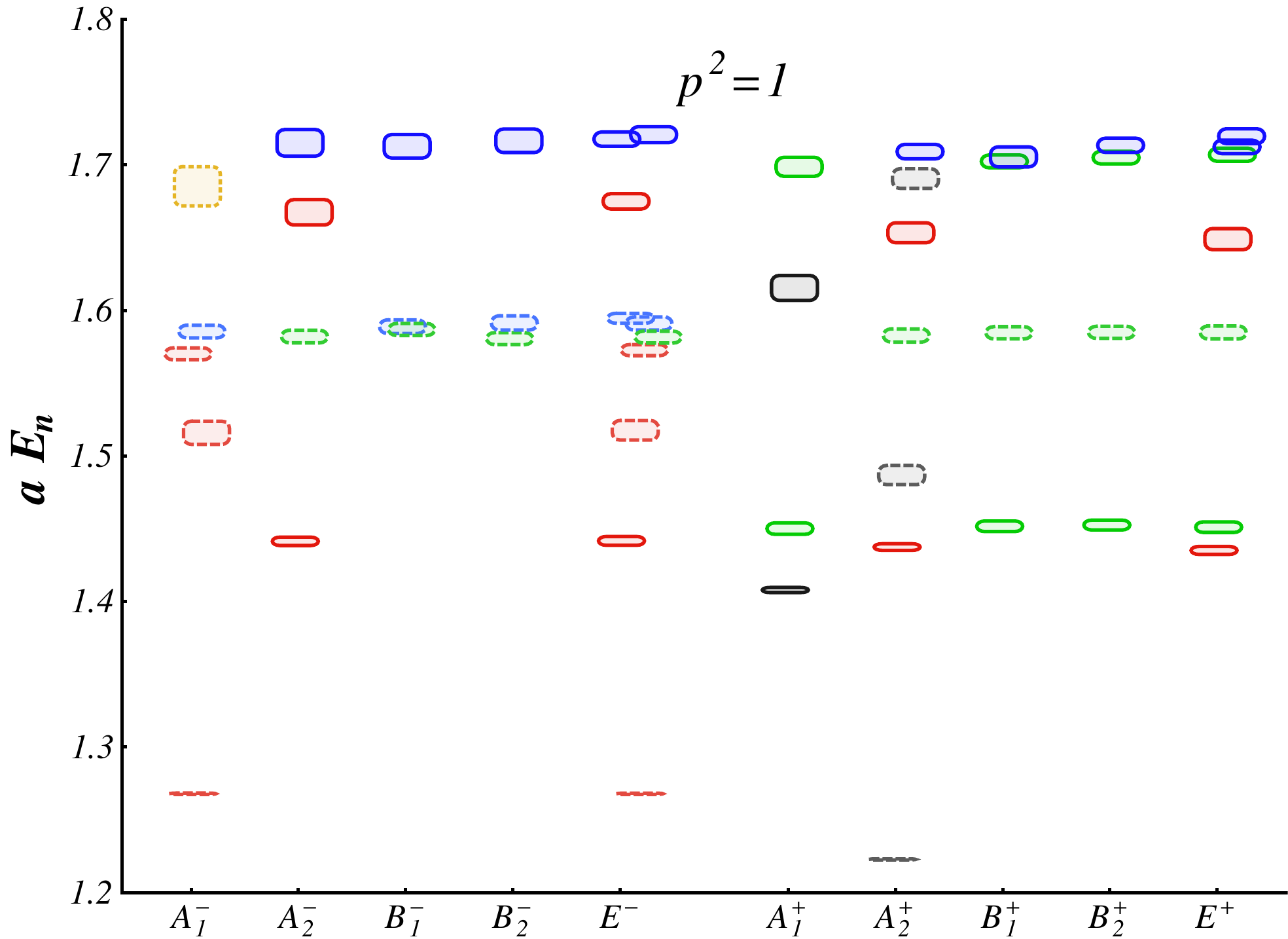}
\includegraphics[scale=0.35]{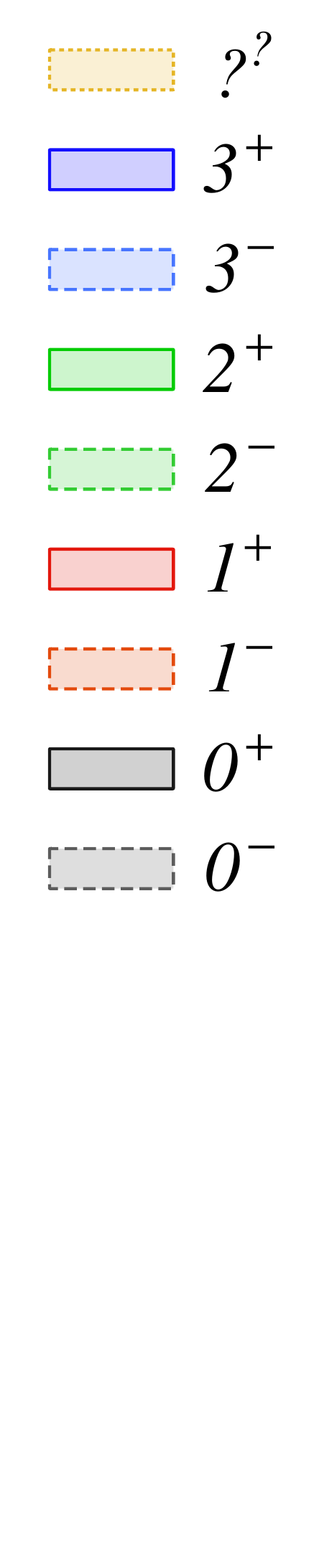}
\end{center}
\caption{$J^P$-identified charmonium spectrum in the moving frame with $\mathbf{p}=(0,0,1)$. 
Irreps $\Lambda^{C}$ of group $Dic_4$ are presented. The colors indicate $J^P$ of states according to the color-coding 
(\ref{eq:colors}). }\label{fig:5-10}
\end{figure*}

\begin{figure*}[th]
\begin{center}
\includegraphics[height=10cm,width=16.2cm]{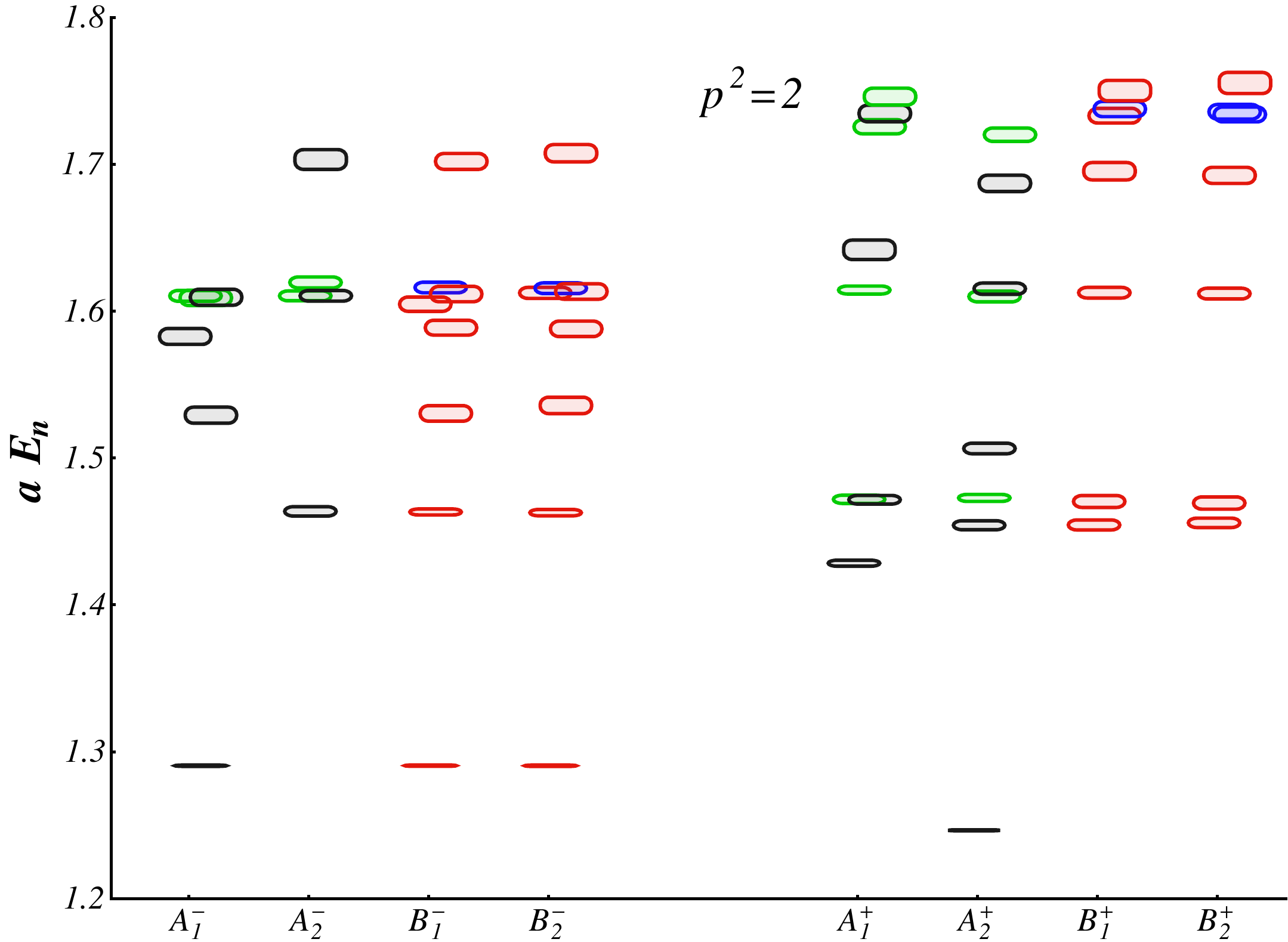}
\includegraphics[scale=0.4]{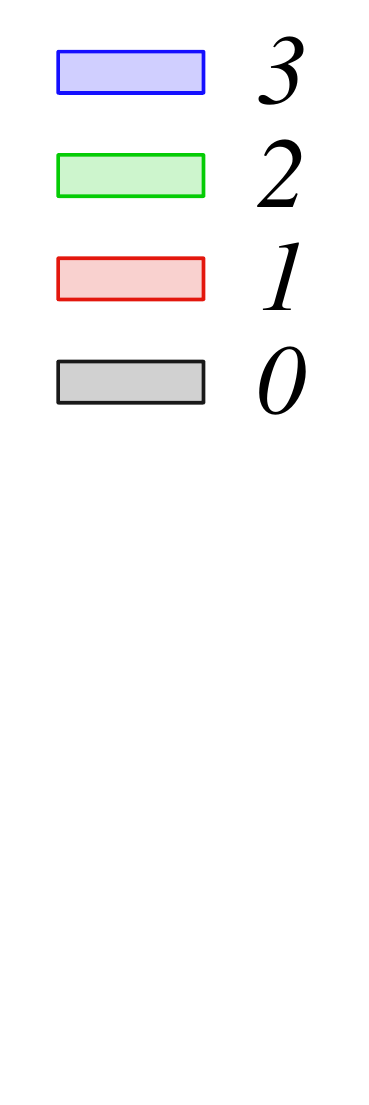}
\end{center}
\caption{$|\lambda|$-identified charmonium spectrum in the moving frame with $\mathbf{p}=(1,1,0)$. All irreps $\Lambda^{C}$ 
of the corresponding little group $Dic_2$ are presented (Table \ref{tab:2-1}). }
\label{fig:5-6}
\begin{center}
\includegraphics[height=10cm,width=16.2cm]{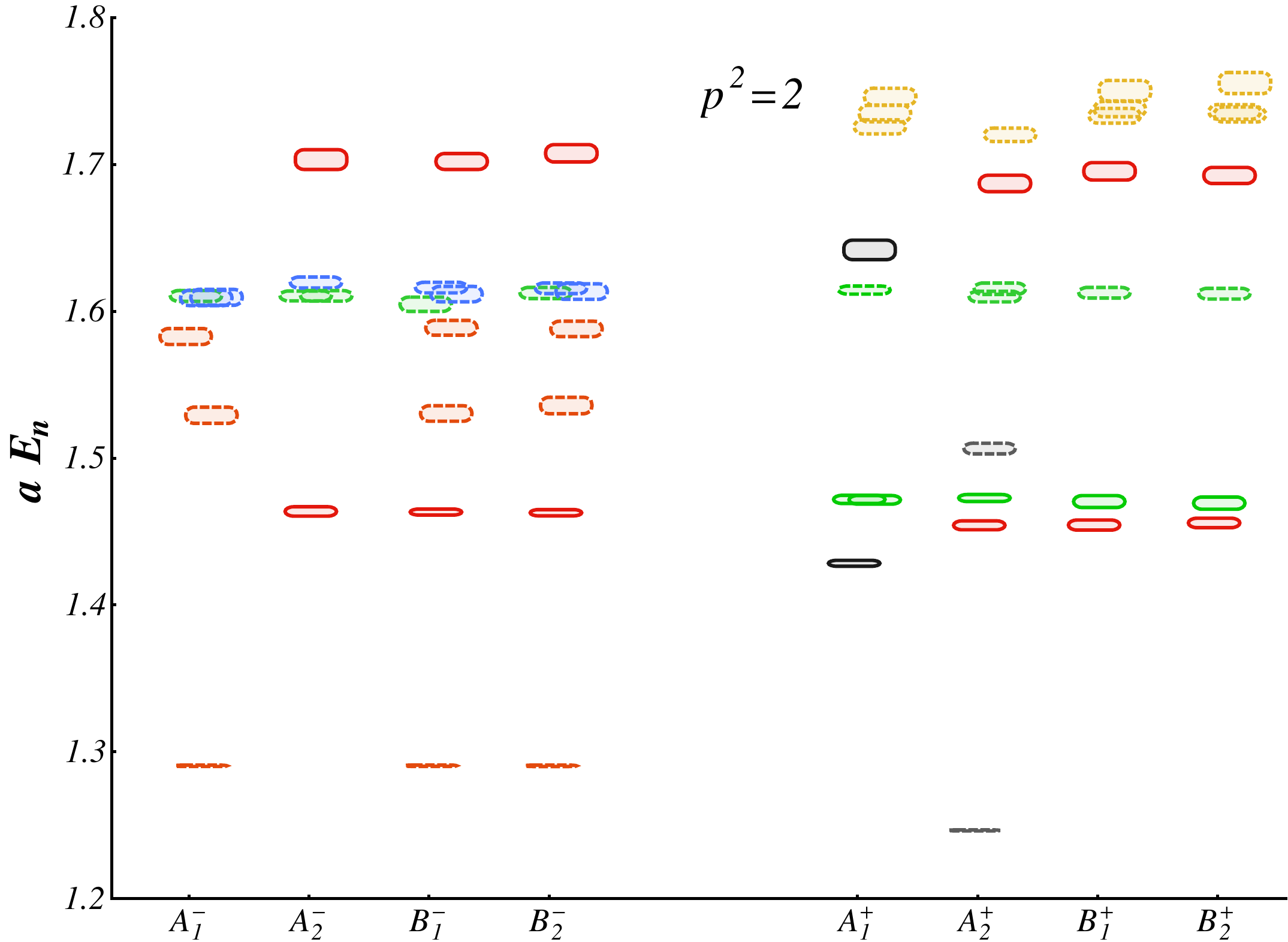}
\includegraphics[scale=0.35]{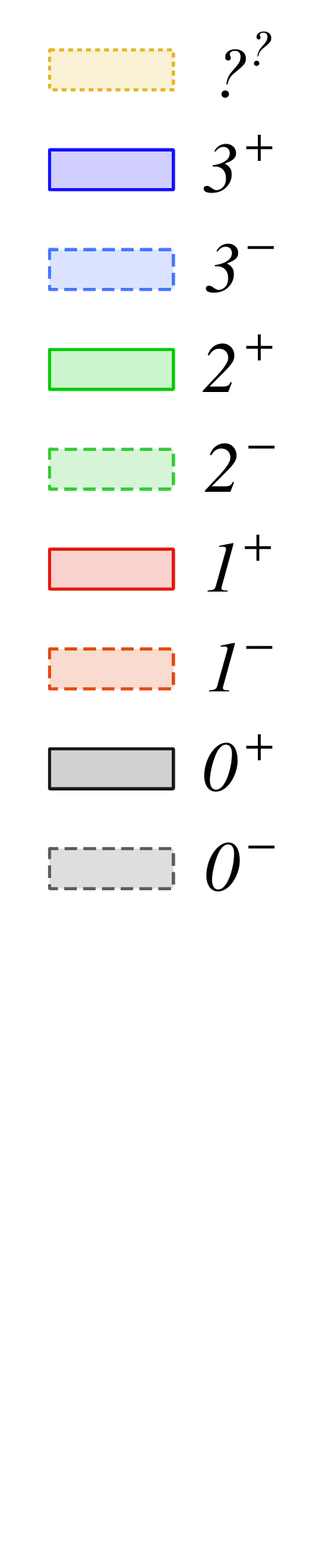}
\end{center}
\caption{$J^P$-identified charmonium spectrum in the moving frame with $\mathbf{p}=(1,1,0)$. Irreps $\Lambda^{C}$ of 
little group $Dic_2$ are presented. }
\label{fig:5-11}
\end{figure*}

In Fig. \ref{fig:5-5} and \ref{fig:5-6}, we present the helicity ($|\lambda|$) assigned charmonium 
spectra in moving frames with momenta ${\bf p}=(0,0,1)$ and ${\bf p}=(1,1,0)$, respectively. The color 
coding of the lattice spectrum indicates the assigned $|\lambda|$ for each extracted level.
 \begin{align}\label{eq:colorsh}
    &\lambda=0   &:& ~\mathrm{black}               \\
    &|\lambda|=1 &:& ~\mathrm{red}      \nonumber  \\
    &|\lambda|=2 &:& ~\mathrm{green}    \nonumber\\
    &|\lambda|=3 &:& ~\mathrm{blue}     \nonumber 
 \end{align}

{\bf Trivial assignments} : Helicity assignments of lattice levels in the spectrum with momentum 
${\bf p}=(0,0,1)$, {\it i.e.} the $Dic_4$ symmetry group, are relatively straightforward. According to 
Table \ref{tab:2-1}, $\lambda=0$ is the only helicity appearing in $A_1$ or $A_2$ among $|\lambda|<4$. 
Thus all levels in the $A_1$ and $A_2$ irreps in $Dic_4$ have helicity 0. Similarly 
$|\lambda|=2$ is the only helicity appearing in the $B_1$ or $B_2$ irreps among $|\lambda|<4$. Hence all 
the levels observed in $B_1$ and $B_2$ can trivially be assigned $|\lambda|=2$. Linear combinations of 
$\lambda=\pm2$ continuum states are expected to appear as pairs of energy levels, degenerate up to 
the effects of the reduced symmetry on the lattice (and finite volume effects), one each in the spectrum of 
$B_1$ and $B_2$. Thus the spectra of these two irreps are expected to be similar, with all levels representing 
$|\lambda|=2$ continuum states. The only non-trivial situation in $Dic_4$ arises in the $E$ irreps, where $|\lambda|=1$ 
as well as 3 can contribute. Disentangling the helicity 1 and 3 levels requires information from the 
$\tilde Z^i_n$ factors as will be discussed shortly.

{\bf Energy degeneracies} : In the charmonium spectrum with momentum ${\bf p}=(1,1,0)$, {\it i.e.} the 
$Dic_2$ symmetry group, each of the four lattice irreps can have contributions from two different 
helicities. Hence there are no trivial helicity assignments possible. According 
to the subduction patterns in Table \ref{tab:2-1} for the $Dic_2$ frame, a $\lambda=0$ state can appear 
only in either $A_1$ or $A_2$, depending on the underlying $J^P$ of the continuum state. 
Linear combinations of the $\lambda=\pm2$ continuum states appear in the spectrum of both $A_1$ and 
$A_2$, degenerate up to the effects of reduced symmetry on the lattice. Hence appearance of 
energy levels in the $A_1$($A_2$) spectrum with no near degenerate partner levels in the $A_2$($A_1$) 
spectrum indicates a $\lambda=0$ assignment. However, appearance of a pair of near degenerate energy 
levels one each in both $A_1$ and $A_2$ could either indicate a possible $|\lambda|=2$ assignment or 
an accidental degeneracy of two $\lambda=0$ levels. A $|\lambda|=2$ assignment for these levels 
implies a possible minimum spin assignment, {\it i.e.}  $J\geq 2$. In this case, one should also find 
another level corresponding to the $\lambda=0$ component nearly degenerate with the $|\lambda|=2$ 
candidates, in either $A_1$ or $A_2$ depending on the $J^P$ of the state in the continuum. Thus 
the levels around $aE_n\sim 1.47$ in the spectrum of $A_1^+$ and $A_2^+$ could be composed of a pair 
of $|\lambda|=2$ levels one each in $A_1^+$ and $A_2^+$ and a $\lambda=0$ level in $A_1^+$. Similarly 
one can make $|\lambda|$ assignments for levels around $aE_n\sim 1.61$ in the spectrum of $A_1^+$ and 
$A_2^+$. The helicity assignment for the levels around $aE_n\sim 1.6$ in the spectrum of $A_1^-$ and 
$A_2^-$ is more complicated and requires some guidance from the spectrum of $B_1^-$ and $B_2^-$ or 
inputs from a study of overlap factors.


Similar to $|\lambda|=2$, linear combinations of states with $\lambda=\pm1$ and $\lambda=\pm3$ appear 
as near degenerate levels in the spectrum of both $B_1$ and $B_2$ in $Dic_2$. Consequently they are 
expected to mimic each other, similar to the case of $B_1$ and $B_2$ irreps in $Dic_4$. However, they 
will be a mixture of $|\lambda|=1$ and $|\lambda|=3$ continuum states. A pair of near degenerate energy 
levels, one each in the spectrum of both $B_1$ and $B_2$, with no other near degenerate levels suggests a 
$|\lambda|=1$ assignment. Thus the lowest three levels in the spectrum of $B_1$ and $B_2$ for both 
$C$-parities can be assigned $|\lambda|=1$, indicating also a possible minimum spin assignment 
$J\geq 1$. A non-zero spin assignment implies the presence of another near degenerate $\lambda=0$ level 
in either $A_1$ or $A_2,$ depending on the $J^P$ of the continuum state. A possible $|\lambda|=3$ assignment 
to any pair of levels in the spectrum of $B_1$ and $B_2$ is indicated by appearance of additional near 
degenerate levels corresponding to the $\lambda=\pm1$ components of the same continuum state at rest, 
similar to what was argued in the case of $|\lambda|=2$. However, such an assignment is complicated as a 
$\lambda=\pm1$ state appears in both $B_1$ and $B_2$, unlike the case of a $\lambda=0$ state, which 
appears only in either $A_1$ or $A_2$. Beyond this point, a study of $Z$ factors becomes crucial. 

A flavor of $J^P$ assignments is already evident from these arguments on the expected degeneracies between 
different helicity components of the same continuum state across different lattice irreps. In the 
discussion above, they are intended to affirm the initial helicity assignments based on 
energy degeneracies. 

\begin{figure}[h]
\includegraphics[width=0.45\textwidth,clip]{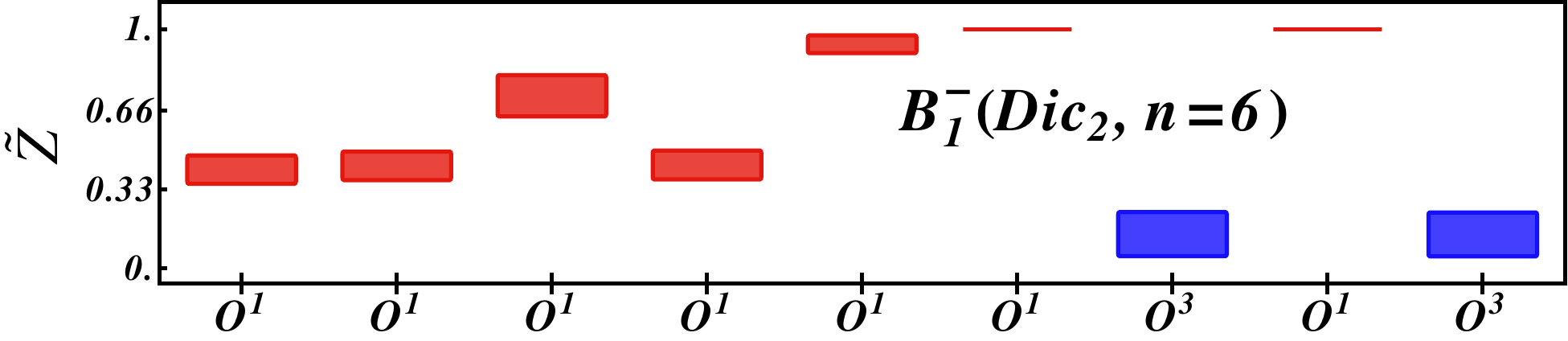}
\includegraphics[width=0.45\textwidth,clip]{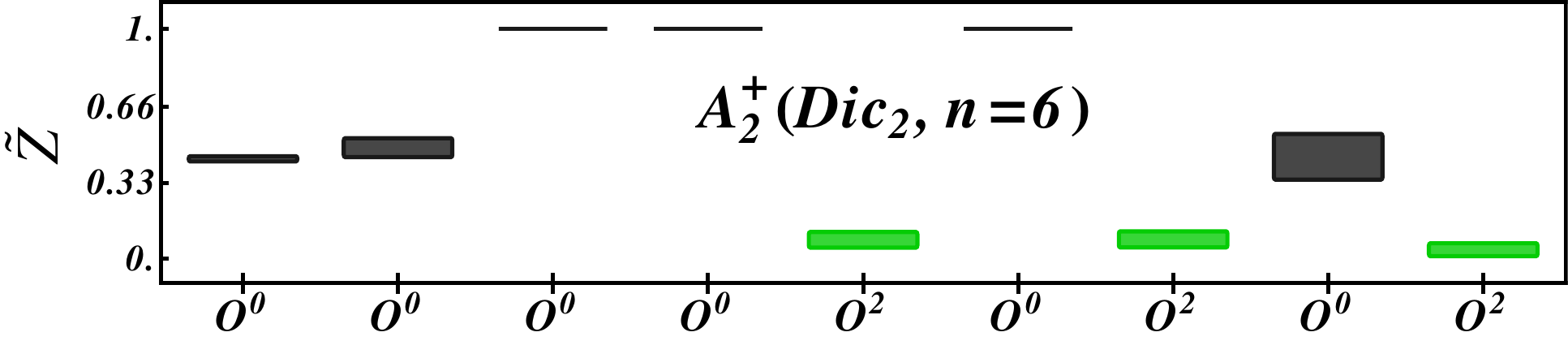}
\caption{$\tilde Z_i^n$ factors for subsets of operators from $B_1^{-}$ and $A_2^+$ of $Dic_2$. This 
figure demonstrates the helicity assignment for the sixth lattice level in these two irreps. The superscripts 
in the $x$-axis labels refer to the helicity of the original continuum interpolator. For brevity, other 
indices are suppressed in the operator labels. The color coding in the data follows Eq. (\ref{eq:colorsh}).}\label{fig:5-7}
\end{figure}

{\bf $\mathbf{Z}$ factor enhancements} : A study of $Z$ factor enhancements on this partly helicity 
assigned spectrum not only verifies initial $|\lambda|$ assignments based on energy degeneracies, 
but also disentangles the helicity composition of all energy levels in the spectrum in a transparent way. 
To demonstrate this, we show $\tilde Z_i^n$ factors for the sixth lattice level for a subset of 
interpolators in the $A_2^+$ and $B_1^-$ irreps in $Dic_2$ in Fig. \ref{fig:5-7}. The levels in the $A_2^+$ 
and $B_1^-$ irreps can clearly be identified to have dominant $\tilde Z$ factors with lattice interpolators, 
$O^{\lambda=0}$ and $O^{|\lambda|=1}$, respectively\footnote{For clarity in the argument, other indices 
are suppressed in the operator labels.}. In this way one can make reliable helicity assignments for all 
the levels in the lattice spectrum. 

\begin{figure}[h]
\includegraphics[height=3cm,width=8.1cm]{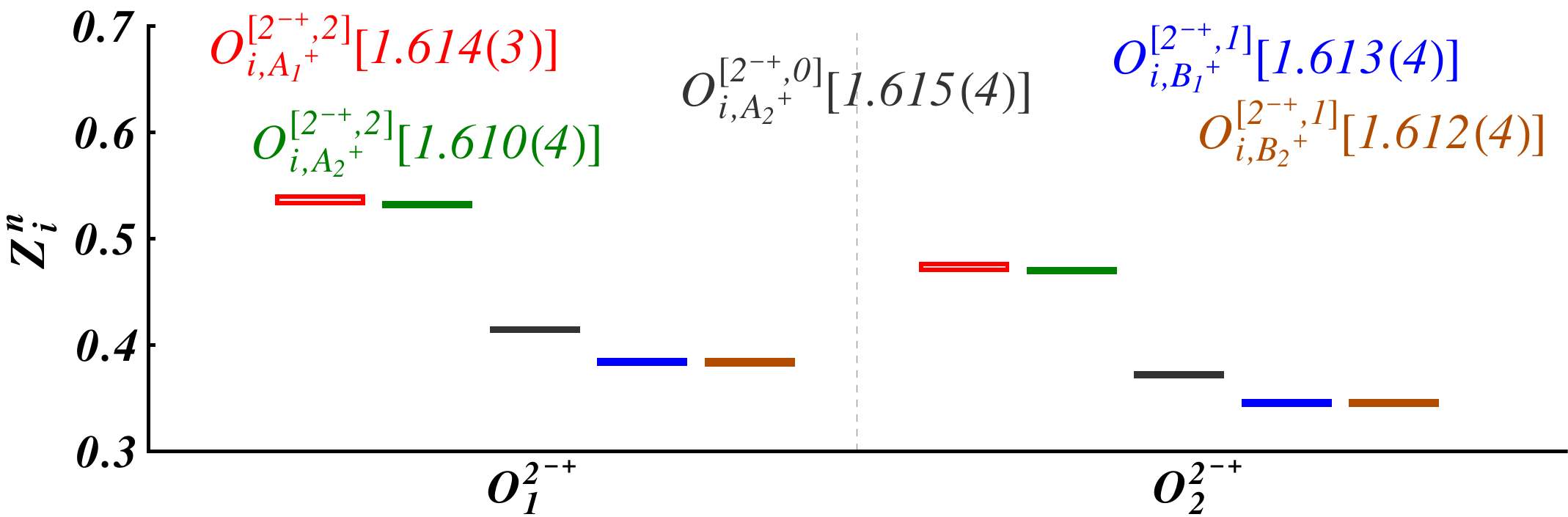}
\caption{A comparison of the $Z_i^n$ factors to confirm the quantum number assignments.
Shown are the $Z$ factors for two continuum interpolators that determine the levels close to $aE_n=1.61$
in all four irreps within $Dic_2$.}\label{fig:5-8}
\end{figure}

{\bf $\mathbf{Z}$ factor degeneracies} : Similar to the study in the rest frame, one can further investigate 
degeneracies in $Z^n_i$ factors of pairs of near degenerate energy levels assigned with the same helicity 
across different lattice irreps (Eq. (\ref{Zmomequal})). In Fig. \ref{fig:5-8}, we compare the $Z_i^n$ factors 
to confirm the quantum number assignment for five levels close to $aE_n=1.61$ in the $C=+$ 
spectrum of all four irreps within $Dic_2$ ({\it c.f.} Fig. \ref{fig:5-6}). Plotted are $Z$ factors of these 
states for lattice interpolators across all lattice irreps originating from two rest frame continuum interpolators, 
$O^{2^{-+}}_{i=1,2}$. As summarized in Table \ref{tab:2-1}, $\lambda=\pm2$ components of $O^{2^{-+}}_{i=1,2}$ 
are subduced to $A_1^+$ and $A_2^+$, while its $\lambda=\pm1$ components are distributed on to $B_1^+$ and 
$B_2^+$. The $\lambda=0$ component of this interpolator falls into the $A_2^+$ irrep, as it has negative parity. 
Since lattice interpolators are built from moving frame continuum operators with good helicity, $Z_i^n$ factors 
for the levels representing the same continuum state $|\lambda|$ across different irreps are expected to be 
degenerate (Eq. (\ref{Zmomequal})). It is evident from the figure that $Z_i^n$ factors of the level in $A_1^+$ and 
one level in $A_2^+$ are degenerate, indicating a $|\lambda|=2$ assignment. Similarly, the degenerate $Z_i^n$ 
factors for the levels in $B_1^+$ and $B_2^+$ indicate a $|\lambda|=1$ assignment. The only remaining level 
in $A_2^+$ with no partners to associate can trivially be assigned $\lambda=0$. Following these procedures, 
we arrive at the charmonium spectrum, as shown in Figs. \ref{fig:5-5} and \ref{fig:5-6}, with 
reliably identified helicities.

\begin{center}
{\it Identifying $J^P$ from known $|\lambda|^{\tilde \eta}$}
\end{center}

Once we have the helicity-identified charmonium spectrum, we can proceed to spin-parity 
assignments, which we perform considering the helicity components and $Z$ factors as 
described below.

{\bf Helicity components } : A meson with continuum $J^P$ can have helicities $|\lambda|\le J$. These 
helicity components are expected to be distributed across the lattice irreps as per Table \ref{tab:2-1}. 
The reduced symmetry of moving frames in the continuum ($U(1)$) does not constrain energies of different 
$|\lambda|$ to be equal. However, constraints from Lorentz symmetry enforce different helicity 
components of a continuum $J^P$ to have the same energy \cite{Thomas:2011rh}. Thus the easiest way to 
make $J^P$ assignments is by investigating the distribution of helicity-identified lattice energy 
levels across the little group irreps (Table \ref{tab:2-1}) and matching them with expected degeneracies. 

First, we discuss the simple case of identifying the isolated $2^{-+}$ state represented by five levels 
close to $aE_n=1.61$ in the $C=+$ spectrum of all four irreps within $Dic_2$ ({\it c.f.} Fig. \ref{fig:5-6}). 
The clear identification of all expected helicity components up to 2 with no remaining energy levels to consider 
and remarkable degeneracy in their energies, as evident from Fig. \ref{fig:5-8}, suggest that all those lattice 
levels to represent the same $J=2$ continuum state. Furthermore, a $J=2$ assignment plus the $\lambda=0$ level 
appearing in $A_2^+$ indicate a $P=-$ assignment for the continuum state in the rest frame. Alternate spin 
assignments are excluded due to the presence of helicity 2 levels. The presence of any additional continuum 
state is excluded as all five levels under consideration are exhausted in describing helicity components of the same 
$J=2$ continuum state. Thus one arrives at a rest frame quantum number $J^{PC} = 2^{-+}$ for the state in the 
$C=+$ $Dic_2$ spectrum around $aE_n\sim 1.61$. 

\begin{figure}[h]
\includegraphics[height=3cm,width=8.1cm]{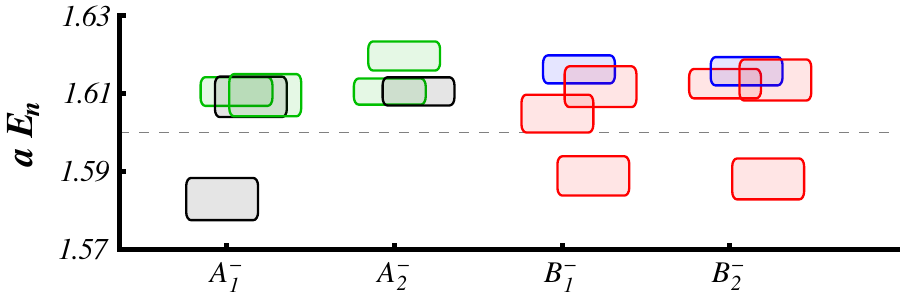}
\caption{Charmonium spectrum in the moving frame with $\mathbf{p}=(1,1,0)$, {\it i.e.} the $Dic_2$ little group, 
in different irreps, $\Lambda^{C}=\Lambda^{-}$ in the energy region $aE_n\in(1.57,~1.63)$. The colors indicate $|\lambda|$ 
of states according to the color-coding in (\ref{eq:colorsh}). }
\label{fig:5-9}
\end{figure}

Now we elaborate on our procedure for a more complicated example of multiple continuum states appearing as a 
band of levels in a narrow energy region. To this end, we consider the energy levels in the region 
$aE_n\in(1.57,~1.63)$ in $Dic_2$ with $C=-$ as highlighted in Fig. \ref{fig:5-9}. With three 
$\lambda=0$ levels collectively in $A_1^-$ and $A_2^-$ irreps, the band must be composed of multiple 
continuum states. The three energy levels well below $aE_n=1.6$ in $A_1^-$, $B_1^-$ and $B_2^-$ can 
be assigned $J^{PC}=1^{--}$. A $J=1$ assignment is evident from the $|\lambda|=1$ levels in the $B_1^-$ 
and $B_2^-$ irreps and the level in $A_1^-$ suggests a $P=-$ assignment. 

There are three energy levels remaining in each of the four irreps. The presence of $|\lambda|=3$ 
levels in $B_1^-$ and $B_2^-$ clearly indicates this energy band should contain a $J=3$ state 
accounting for seven lattice levels : \\
\hspace*{0.5cm} - two $|\lambda|=3$ levels in $B_1^-$ and $B_2^-$, \\
\hspace*{0.5cm} - two $|\lambda|=2$ levels in $A_1^-$ and $A_2^-$, \\
\hspace*{0.5cm} - two $|\lambda|=1$ levels in $B_1^-$ and $B_2^-$ and \\
\hspace*{0.5cm} - a $\lambda=0$ level in either $A_1^-$ or $A_2^-$. \\ 
Of the remaining five levels, the presence of two $|\lambda|=2$ levels in $A_1^-$ and $A_2^-$ 
indicates a $J=2$ state : \\ 
\hspace*{0.5cm} - two $|\lambda|=2$ levels in $A_1^-$ and $A_2^-$, \\
\hspace*{0.5cm} - two $|\lambda|=1$ levels in $B_1^-$ and $B_2^-$ and \\ 
\hspace*{0.5cm} - a $\lambda=0$ level in either in $A_1^-$ or $A_2^-$. \\
Other $J-$assignments are trivially excluded as $J=2$ and $J=3$ account for all the levels 
considered, due to the largest helicity components. 

The only remaining quantum number of these two states to be inferred is parity, which is determined by 
$\tilde \eta = P(-1)^J$ as discussed earlier. However, this is not immediate as it is not evident which of 
the $\lambda=0$ levels in $A_1^-$ and $A_2^-$ corresponds to the $J=2$ and $J=3$ continuum state. 
Nevertheless, an immediate inference from these two $\lambda=0$ levels is that both the $J=2$ and $J=3$ 
states should have the same parity. This is independent of the final spin assignments of these $\lambda=0$ levels.
This means if the $\lambda=0$ level in $A_1^-$ is related to the $J=2$ continuum state, 
then both $J=2$ and $J=3$ states have positive parity. If instead the $\lambda=0$ level in $A_1^-$ is 
related to the $J=3$ continuum state, then both of them have negative parity. In other words, the 
continuum $J^{PC}$ composition of energy levels in this band can be either ($2^{+-},3^{+-}$) or ($2^{--},3^{--}$). 
However, due to the exotic nature of $J^{PC}=2^{+-}$, a $PC=+-$ assignment is less plausible at this energy range. 
Hence, we make a ($2^{--},3^{--}$) assignment for these levels. For the levels identified 
with $\lambda=0$ and $|\lambda|=3$ the spin assignment is complete, whereas for the levels identified 
with $|\lambda|=1$ and $|\lambda|=2$ the spin assignment (between $J=2$ or $J=3$) requires additional 
information contained in the $\tilde Z$ factors, along the lines described in the next subsection. 

\begin{figure*}[bht]
\begin{center}
\includegraphics[height=1.8cm,width=5.7cm]{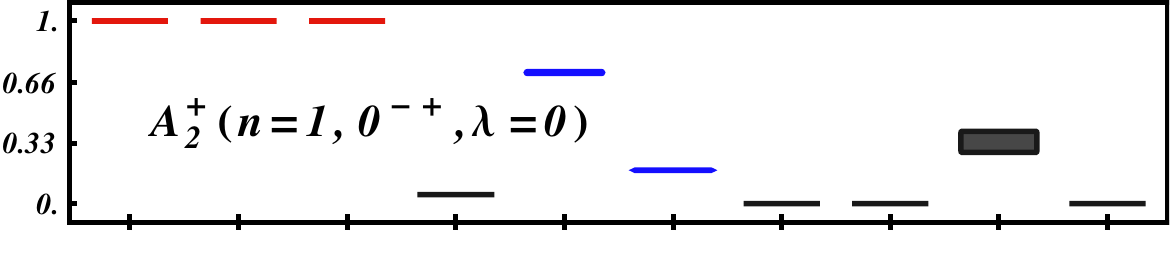}
\includegraphics[height=1.8cm,width=5.7cm]{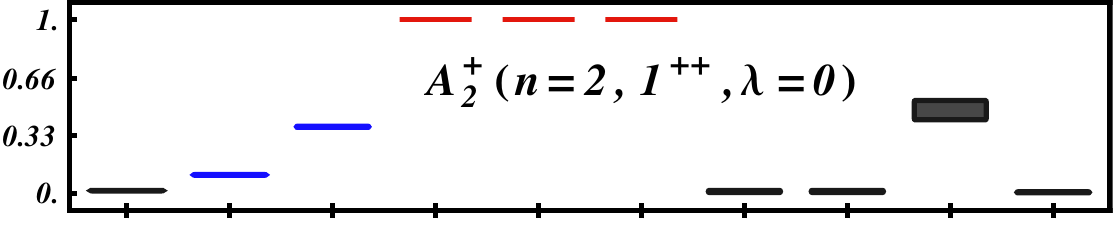}
\includegraphics[height=1.8cm,width=5.7cm]{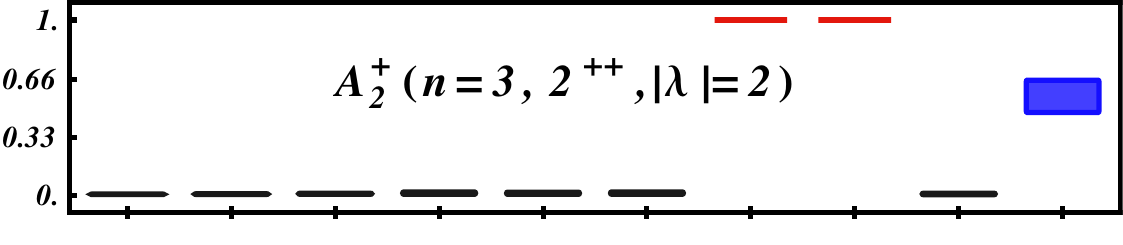}\\
\includegraphics[height=1.8cm,width=5.7cm]{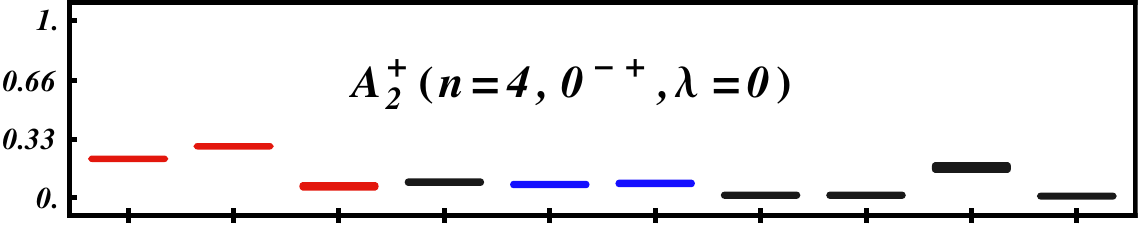}
\includegraphics[height=1.8cm,width=5.7cm]{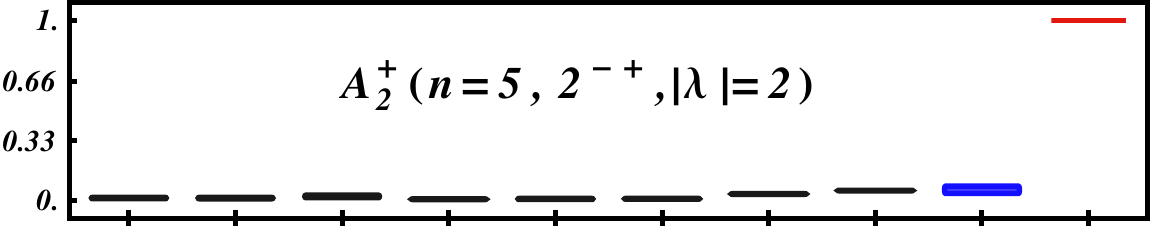}
\includegraphics[height=1.8cm, width=5.7cm]{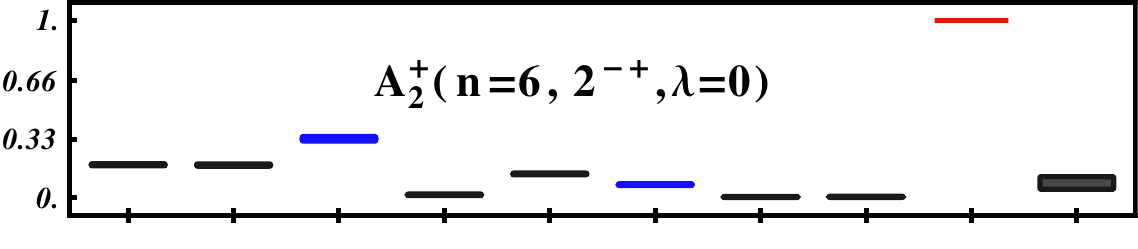} 
\end{center}
\caption{Overlap factors $\tilde Z_i^n$ of the lowest six levels in the $A_2^+$ irrep of $Dic_2$ for operators constructed 
with up to one derivative. The assigned $J^P$ for each eigenstate are provided in the respective pane. Red 
indicates $\tilde Z_i^n$ for interpolators $ O^{[J^{P+},|\lambda|]}_{i,A_2^+}$ that couple to the 
assigned quantum numbers $J^P$ at zero momentum in the continuum. Blue indicates $\tilde Z_i^n$ for interpolators 
that can couple to the assigned quantum numbers only at non-zero momentum in the continuum. Those that cannot 
couple with the assigned quantum  numbers in the continuum are shown in black.}\label{fig:5-13}
\end{figure*}

{\bf Information from $\tilde Z$ factors:}  As mentioned in Section \ref{sec:basicsinflight}, at rest the continuum operator
$O^{J^{P},|\lambda|}_{i}({\mathbf p})$ can only overlap with $|\mathbf{p},J^{PC},\lambda\rangle$. At 
non-zero momentum, the $O(3)$ symmetry group is reduced to the $U(1)$ subgroup and this interpolator can 
overlap with states of different spin and parity ($J'^{P'}$). If the employed momentum 
and the effects of breaking of continuum symmetry on the lattice are small, the $\tilde Z$ factors for 
$O^{[J^{P},|\lambda|]}_{i,\Lambda^C}({\mathbf p})$ are expected to be dominant for states with quantum numbers 
$J^P$ at rest, {\it i.e.}
\begin{equation}
\langle O^{[J^{PC},|\lambda|]}_{i,\Lambda^C}|\mathbf{p},J^{PC},\lambda\rangle >  \langle O^{[J^{PC},|\lambda|]}_{i,\Lambda^C}|\mathbf{p},J'^{P' C},\lambda\rangle  
\label{LSCeq1}
\end{equation}
for $J'^{P'} \not = J^P$. To illustrate this, we present the $\tilde Z$ factors for the lowest 
six levels in $A_2^+$ of $Dic_2$ and a subset of interpolators in Fig. \ref{fig:5-13}. The assigned $J^P$ for 
each level are provided in each pane and the color coding for different operators depends on the $J^P$ of 
the state in the continuum. Red indicates $\tilde Z$ factors for operators $ O^{[J^{P+},|\lambda|]}_{i,A_2^+}$
that couple to the assigned quantum numbers $J^P$ at zero momentum in the continuum. 
Blue indicates $\tilde Z$ factors for operators $O^{[J'^{P'+},|\lambda|]}_{i,A_2^+}$ that can overlap with 
states of quantum numbers $J^P$ in the continuum at non-zero momentum. This information can be deduced from Tables 
X-XV in Ref. \cite{Thomas:2011rh} and the corresponding overlaps are proportional to ${\mathbf p}$  or to
higher powers. Black is used for $\tilde Z$ factors of operators that cannot overlap with quantum 
numbers $J^P$ in the continuum either at rest or with non-zero momentum. Hence the $\tilde Z$ factors in red 
are expected to be dominant in comparison with those in blue and black. This is evident from the figure 
\footnote{ The $\tilde Z$ is not significantly enhanced for the $n=4$ state in $A_2^+$ of $Dic_2$. But its 
spin-parity $J^P=0^-$ is clear, given the absence of near degenerate levels in other irreps ({\it c.f.} Fig. \ref{fig:5-6}).}. 
The large $\tilde Z$ factors in red unambiguously confirm the $J^P$ assignments for all the lattice levels we 
have extracted. Note that the information in Table X-XV of Ref. \cite{Thomas:2011rh}, which helps to differentiate the 
operators into blue and black, was not essential for identifying $J^P$ in Fig. \ref{fig:5-10} and Fig. \ref{fig:5-11}. In this way, we confirm all previously assigned $J^P$ 
and identify the $J^P$ for all the remaining ambiguous levels. Thus taking information from the expected patterns 
of the energy spectrum in moving frames and from the $\tilde Z$ factors into account, we arrive at the $J^{PC}$ 
assigned lattice spectrum in the moving frames, as shown in Figs. \ref{fig:5-10} and \ref{fig:5-11}. The color 
coding is given in Eq. (\ref{eq:colors}). 

We remark that the $Z$ factors for different helicities of a continuum state with non-zero momentum are 
not constrained to be degenerate by the $U(1)$ symmetry group. We indeed find statistically different 
overlaps for different $|\lambda|$ ({\it c.f.} Fig. \ref{fig:5-8}). Hence a comparative study such as in 
Fig. \ref{fig:5-3} is not useful in making the continuum spin assignments in a moving frame.  

\begin{figure}[h]
\includegraphics[height=6.3cm,width=8.7cm]{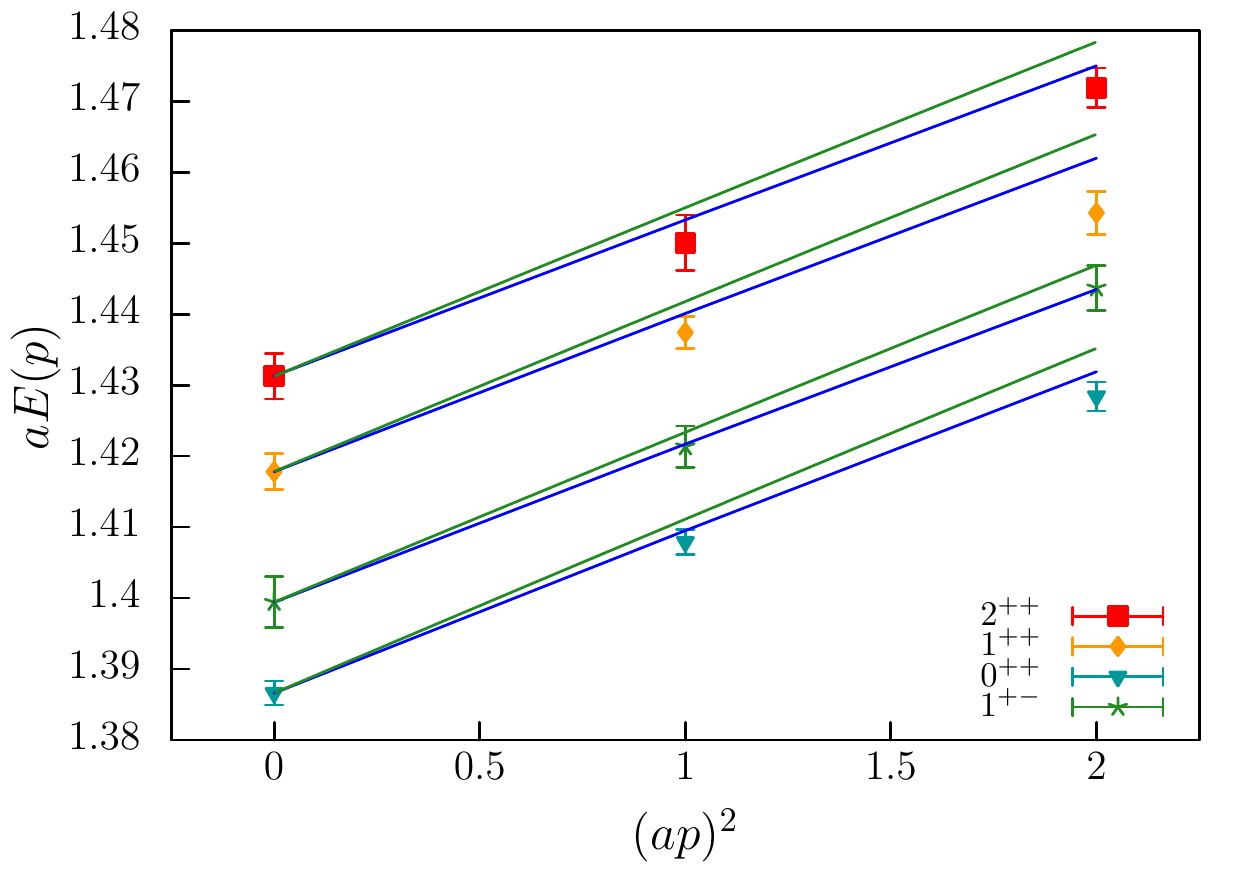}
\caption{The lattice energies at various ${\mathbf p}^2$ for the $1P$ charmonium levels together with the dispersion 
relation (\ref{eq:5-disp}) at integer ${\mathbf p}^2$ joined by lines (blue). The parameter $b$ in the dispersion relation 
is fitted from the $1^{+-}$ charmonium  and the same parameter is employed for the other states. The expectations 
from the continuum dispersion relation are presented as green curves. The points and curves corresponding to $1^{+-}$ 
charmonium are shifted down by a vertical offset, such that its mass is $aE(0)=1.4$, for clarity in the plots.}
\label{fig:5-12}
\end{figure}

{\bf Consistency check using the dispersion relation:} Finally, as a consistency check, we verify that the energies 
of the spin identified spectra from different inertial frames roughly follow a dispersion relation $E(\mathbf{p})$. 
The charmed hadrons at our lattice spacing $a$ are not expected to satisfy the continuum dispersion relation exactly. 
So we test the extracted energies based on a lattice dispersion relation
\begin{equation}
\cosh(bE(\mathbf{p}))=\cosh(bM)+\sum_{i=x,y,z}\biggl[2 \sin\bigl(\frac{b p_i}{2}\bigr)\biggr]^2,\quad \label{eq:5-disp}
\end{equation}
where $M=E(0)$ is the energy extracted at rest and the parameter $b$ can be viewed as an effective lattice spacing 
relevant for this momentum dependence. This dispersion relation interpolates between the continuum relation 
$E(\mathbf{p})=\sqrt{M^2+ \mathbf{p}^2}$ for $b\to 0$ and the  free-boson dispersion relation for $b=a$. 
Figure \ref{fig:5-12} shows an example (blue curves), where the parameter $b\simeq 0.46\; a$ is fit from 
$J^{PC}=1^{+-}$ ($\chi^2$/d.o.f = 0.008) and the same parameter is employed for other $1P$ charmonia, indicating 
that (\ref{eq:5-disp}) results in a reasonable momentum dependence for those. The green curves present 
expectations using the continuum dispersion relation ($b\to 0$), which indicate the inferences from consistency 
checks using different dispersion relations remain intact. The above conclusions are also found to be robust
with an equivalent consistency check using the Fermilab dispersion relation for heavy quarks \cite{ElKhadra:1996mp}. 

\begin{figure*}[th]
\begin{center}
\includegraphics[height=10cm,width=17.1cm]{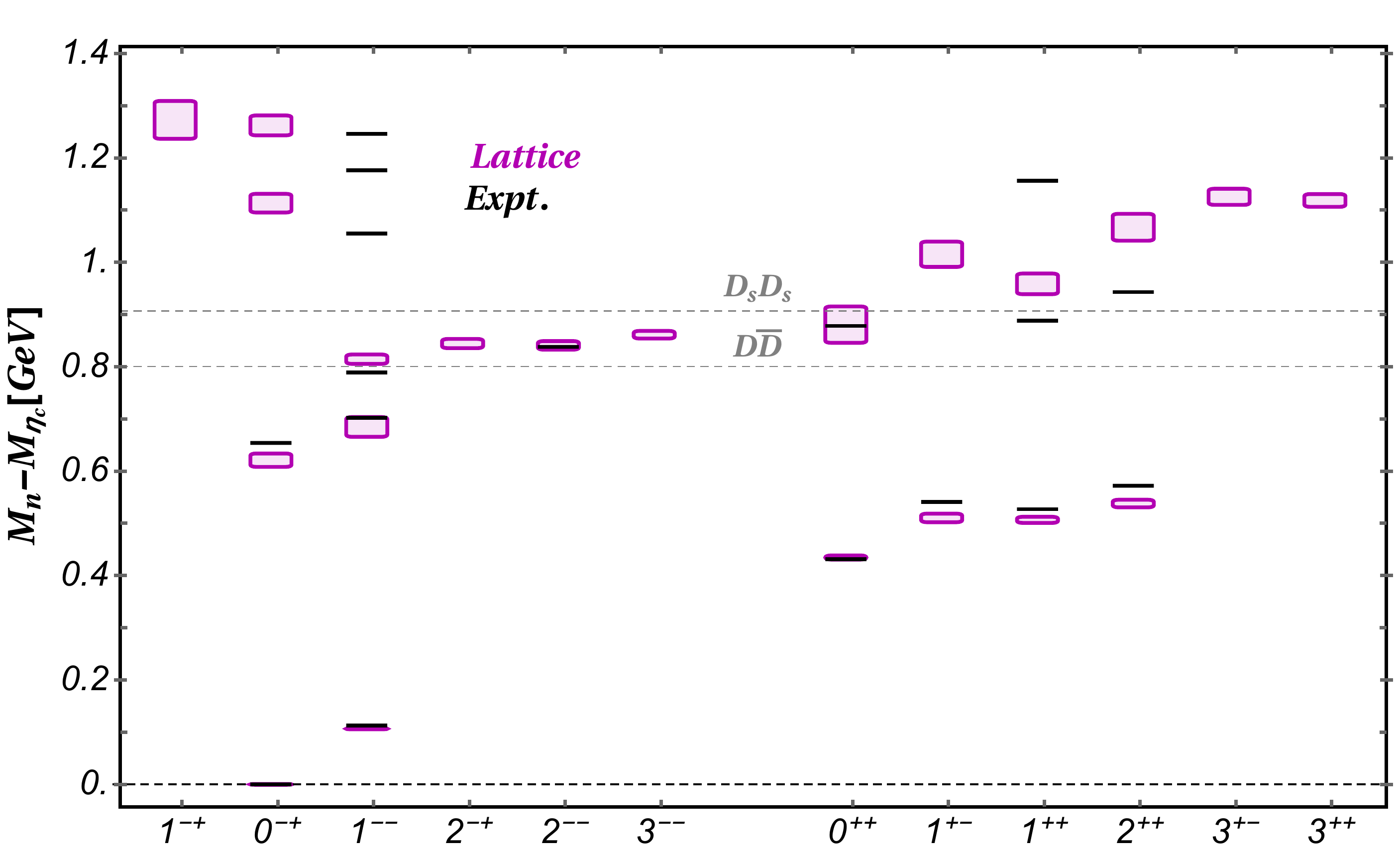}
\end{center}
\caption{ Summary of the charmonium masses extracted from our lattice
  simulation for various $J^{PC}$ within the single-hadron approach
  (using only quark-antiquark interpolators) (magenta). 
  Mass splittings from the mass of the $\eta_c$ are shown 
  together with the experimental values  \cite{Tanabashi:2018oca} (black). We
  emphasize that these results are a qualitative illustration from a single lattice ensemble at
  unphysical pion mass, and likely  do not capture mesons close to decay
  thresholds or broad resonances correctly. The candidate for the first 
  excited scalar charmonia from experiment is not yet settled and we show 
  the mass of the recently discovered $X(3860)$ \cite{Chilikin:2017evr}. 
  Dashed lines denote the location of the lowest open-charm thresholds on 
  our lattice. The reference mass is $M_{\eta_c}^{exp}\simeq 2.984~GeV$ in experiment.   }
\label{fig:5-14}
\end{figure*}

\begin{figure*}[th]
\begin{center}
\includegraphics[height=10cm,width=17.1cm]{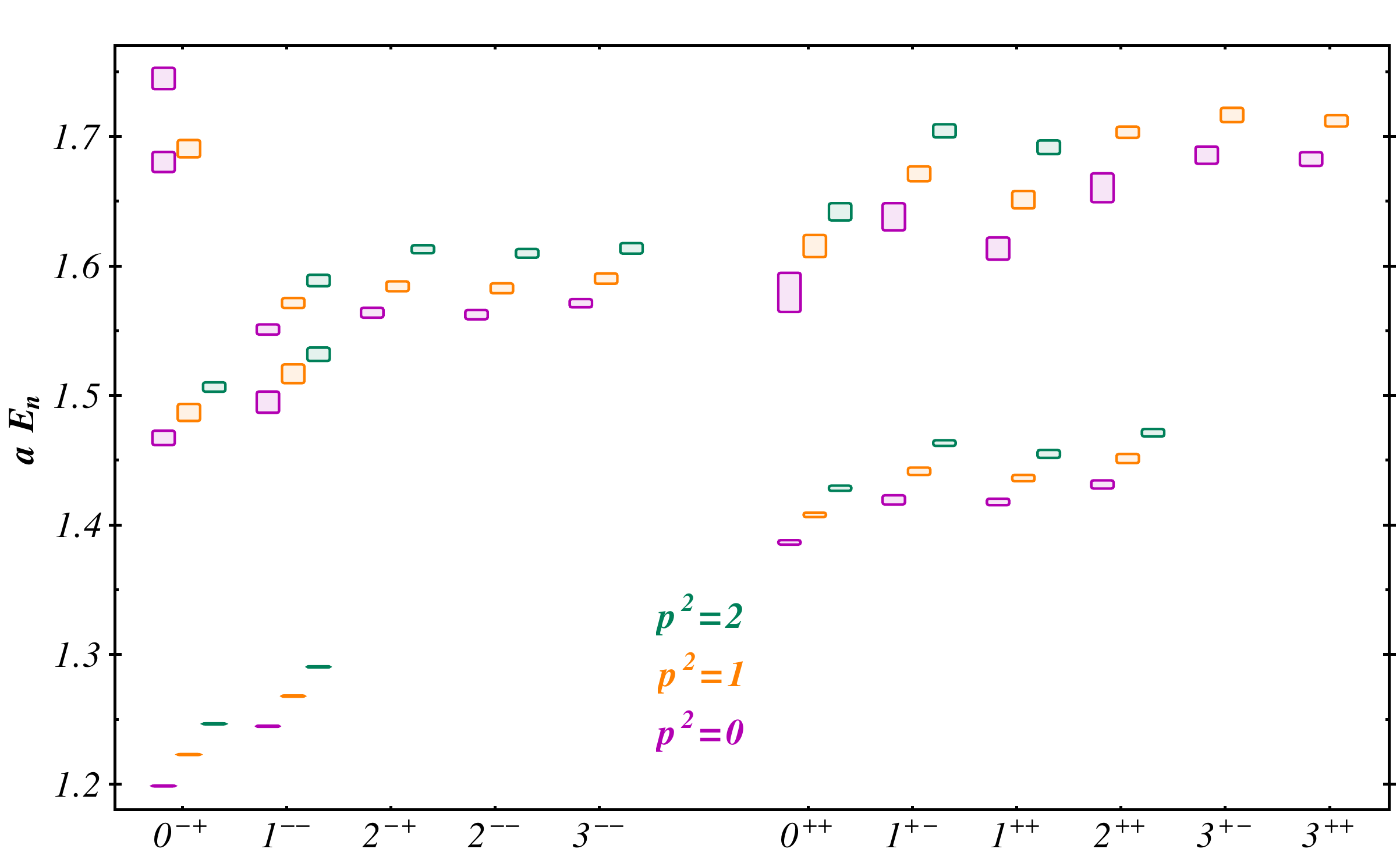}
\end{center}
\caption{Summary plot of the charmonium energies $Ea$ (in lattice units) for various $J^{PC}$, compared for  three values of the momenta ${\mathbf p}^2$ (in units $(2\pi/L)^2$) . Shown are all  states with   reliably-identified spin $J$ and parity $P$, which are found at rest and at least for one non-zero momentum frame. }
\label{fig:5-15}
\end{figure*}

\subsection{ Charmonium summary  for various $J^P$}\label{sec:results-summary}

The $J^{PC}$-identified charmonium spectra with various momenta within the single hadron approach are 
summarized in Figs. \ref{fig:5-14} and \ref{fig:5-15}. Only the states with reliably identified spin 
and parity are shown and the energy values are based on Figs. \ref{fig:5-1}, \ref{fig:5-10} and \ref{fig:5-11}
\footnote{Eigenstates with certain $J^{PC}$ are present in several lattice irreps and we choose 
the average of energies from all the irreps for each jackknife
sample. Note that the levels in different irreps receive different shifts
  from the finite volume and the discretization used. Averaging those values
  provides a spectrum suitable for a qualitative comparison to experiment
  rather than a quantitative prediction.}. 

Figure \ref{fig:5-14} shows the charmonium mass spectrum determined at rest in terms of mass 
splittings with respect to the mass of the $\eta_c$ meson. These mass splittings are compared to 
those from experiments \cite{Tanabashi:2018oca}. Except for $1^{-+}$, the extracted levels are 
likely candidates for conventional quark model charmonia $\bar cc$. The $1^{-+}$ state is a 
candidate for the hybrid charmonium $\bar cGc$ with exotic $J^{PC}$, {\it i.e.} quantum numbers 
that cannot result from a quark-antiquark configuration. The features of the lattice and 
experimental spectrum agree qualitatively, while a quantitative agreement is not expected due 
to several shortcomings in our calculation. First of all, we utilize a lattice QCD ensemble 
with heavier than physical pion mass ($m_\pi\simeq 280~$MeV) and our charm quark mass corresponds 
to a $D$ meson mass about $80~$ MeV lower than its physical value, as discussed in Section 
\ref{sec:setup}. This work ignores the resonant nature of charmonia above the thresholds (indicated 
for $D\bar D$ and $D_sD_s$ by dashed lines) and the resulting mass estimates are naively expected to be correct to the order 
of the respective decay widths. We also ignore the effects of thresholds on the near-threshold states. 
This renders the first excited state $1^{++}$ state too high with respect to experimental $X(3872)$, 
which is common to all lattice results within the single hadron approach. The features of the extracted 
lattice spectra in Fig. \ref{fig:5-14} also roughly agree with other lattice results based on the single 
hadron approach (all in the rest frame) aimed at excited $\bar cc$  
\cite{Liu:2012ze,Cheung:2016bym,Dudek:2007wv,Bali:2011rd,Mohler:2012na}. 

Now we turn to the spectra with non-zero momenta, as the identification of their spin and parity was 
the main purpose of our work. The $J^P$ assigned energy spectra of charmonium in different inertial frames  
are compared in Fig. \ref{fig:5-15}. The energies increase with the momentum roughly as expected by 
the dispersion relation (Eq. (\ref{eq:5-disp})); an example of a more detailed comparison was shown in Fig. 
\ref{fig:5-12}. One does not expect the dispersion relation $E(\mathbf{p})$ to be respected exactly 
for states near and above thresholds, since energies at different momenta receive different shifts, depending on 
the positions of the discrete non-interacting two-meson states. This plot presents only the 
spectral levels with reliably identified $J^P$ and reflects the fact that this identification gets more 
challenging as momentum increases due to the reduction of the discrete symmetries on the lattice 
(Table \ref{tab:2-1}) as well as due to the degradation of signal-to-noise at larger energies. 
Therefore, some of the higher-lying charmonia could be reliably assigned $J^P$ only for the 
lower momenta.

\section{Conclusions}\label{sec:conclusions}

This paper studies the spectra and spin-parities of  charmonia at rest and in flight. 
The analysis is performed using the distillation method on a single $N_f=2+1$ CLS ensemble 
with $m_\pi\simeq 280~$MeV and a somewhat smaller than physical charm quark mass.

In the continuum, the spin $J$ and parity $P$ of a hadron are good quantum numbers in 
its rest frame, while the helicity $\lambda$ is a good quantum number for a particle 
in flight. In experiment, the $J^P$ of a hadron in flight is determined by making a 
Lorentz transformation of its decay products to its rest frame. In lattice QCD, there is no 
direct analogue of this procedure due to the reduced Lorentz symmetry. 

We show a reliable procedure to determine the underlying rest-frame $J^P$ of charmonia in 
flight within lattice QCD. To identify $J^P$ of a certain eigenstate, it is not enough 
to consider this state in isolation. One should consider most of the eigenstates up to 
the desired energy in all lattice irreducible representations. The challenge is that 
eigenstates with many different $J^P$ contribute to a given irreducible representation. 
The strategy is to determine the energies of all these eigenstates and their overlaps to 
lattice versions of carefully-constructed operators $O^{[J^P,|\lambda|]}(p)\simeq \bar cc$ 
\cite{Dudek:2010wm,Thomas:2011rh}, which in the continuum couple only to charmonia with 
spin-parity $J^P$ at $p=0$ and to states with helicity $\lambda$ at $p\not =0$. The $J^P$ 
are determined by considering degeneracies of energies or overlaps across different 
irreducible representations, and by considering the sizes of appropriately normalized 
overlaps. In this way, we reliably identify the corresponding spin and parities for all 
ground and excited charmonia with masses $m\leq 4.0$ GeV, $J\leq 3$ and $p^2\leq 2\;(2\pi/L)^2$. 

It is straightforward to apply the procedure illustrated in this work to study the excited 
hadron spectrum on any other lattice QCD ensemble. The relevance of different interpolators 
could change depending on the system (light mesons, heavy mesons or heavy quarkonium) under investigation. 
Naively one may have to repeat the whole analysis for each system being studied and on each ensemble. 
Even if this is the case, for heavy quarkonium spectroscopy qualitative information on 
the operator state overlaps from this investigation can 
alleviate the efforts to extract the spin identified spectrum on a different ensemble\footnote{The 
energies and $Z$ factors can be made available on request.}. Note that the energy ordering of near 
degenerate states can vary depending on the parameters of the ensemble. Hence one cannot simply carry 
over the $J^P$ assignments from one ensemble to the next based on energy ordering alone. 

The results in this paper will be valuable for our ongoing study of charmonium resonances in (coupled-channel) 
meson-meson scattering, where we combine the quark-antiquark basis used in this study with 
meson-meson interpolators to go beyond the single hadron approach.

 \acknowledgments

 We would like to thank G. Bali and C. Thomas for valuable discussions. We would like to thank 
 C. B. Lang for helping with the variational analysis codes. The Regensburg group was supported 
 by the Deutsche Forschungsgemeinschaft Grant No. SFB/TRR 55. S. P.  was supported by Slovenian 
 Research Agency ARRS (research core funding No. P1-0035 and No. J1-8137). M. P. acknowledges 
 support from the EU under grant no. MSCA-IF-EF-ST-744659 (XQCDBaryons). We are grateful to 
 the Mainz Institute for Theoretical Physics (MITP) for its hospitality and its partial support 
 during the completion of this work. We thank our colleagues in CLS for the joint effort in the 
 generation of the gauge field ensembles which form a basis for the here described computation. 
 The simulations were performed on the Regensburg iDataCool and Athene2 clusters and the SFB/TRR 
 55 QPACE2 \cite{Arts:2015jia} and QPACE3 machines. Part of the simulations were preformed at 
 the local cluster at the Department of Theoretical physics at Jozef Stefan Institute. The 
 CHROMA~\cite{Edwards:2004sx} software package was used extensively along with the multigrid 
 solver implementation of Refs.~\cite{Heybrock:2014iga,Heybrock:2015kpy,Richtmann:2016kcq,Georg:2017diz}~(see also Ref.~\cite{Frommer:2013fsa}).

\appendix

\section{Energies $E_n$ and operator state overlaps $Z_i^n$}
\label{app:fits}

\begin{figure*}[tb]
\begin{center}
\includegraphics[width=0.6\textwidth]{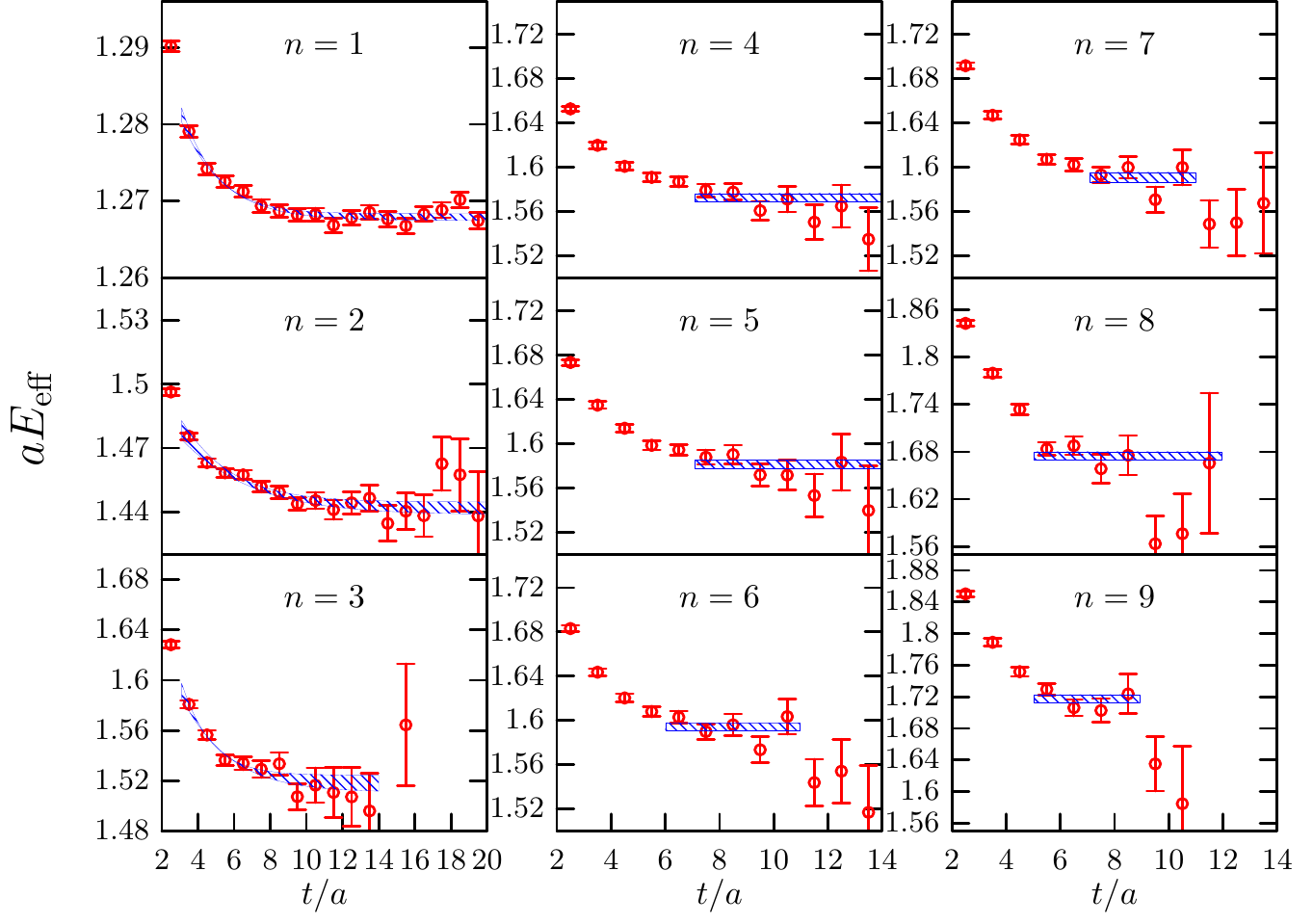}
\end{center}
\caption{Effective masses~($aE_{\rm eff}$) of the first nine eigenvalues of the
  generalized eigenvalue problem for the $E^-$ representation of
  $Dic_4$. The blue shaded regions indicate the 1~(2) exponential fits to
eigenvalues $4-9$~($1-3$) and the corresponding uncertainty. The time extent of
the shaded region shows the range chosen for the fit.  }\label{fig:app-2}
\end{figure*}

\begin{figure*}[tb]
\begin{center}
\includegraphics[width=17cm,height=8cm]{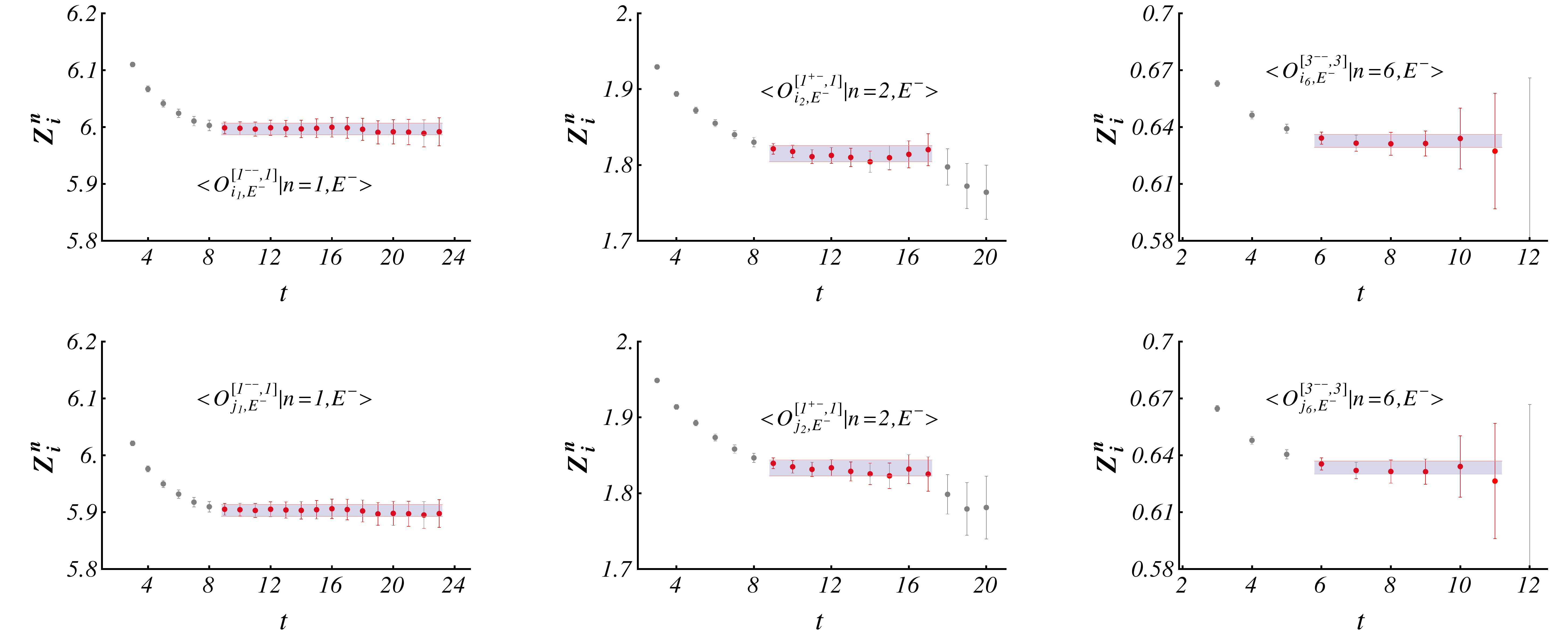}
\end{center}
\caption{$Z$ factors for a selected pair of interpolators separately for $n=1$, 2 and 6
excitations in the $E^-$ irrep of $Dic_4$ frame.
The red shaded regions indicate the constant fits to the plateau in the $Z$ factors and the corresponding errors.
The red points are those included in the fits. }\label{fig:app-3}
\end{figure*}

Fig.~\ref{fig:app-2} displays the effective masses plotted against the energy fit estimates
for the example of the $E^-$ irrep of $Dic_4$. Fig. \ref{fig:app-3} shows the
$Z$ factors for a selected pair of interpolators for $n=1$, 2 and 6 excitations in the same irrep.

\bibliography{references}

\end{document}